\documentclass[aps,pra,twocolumn,superscriptaddress,showpacs,amsfonts,amsmath,floatfix]{revtex4-2}

\usepackage{amsmath, amsfonts, amssymb, amsthm, bbm}
\usepackage{dsfont}
\usepackage[T1]{fontenc}
\usepackage{enumitem}
\usepackage{bm}
\usepackage{mathtools}
\usepackage{graphicx}
\usepackage{epstopdf} 
\usepackage{epsfig}
\usepackage[normalem]{ulem}
\usepackage{dcolumn}
\usepackage[table,xcdraw,svgnames]{xcolor}
\usepackage{bm}
\usepackage[breaklinks=true,
        ]{hyperref}  
\usepackage{breakurl}
\usepackage{times}
\usepackage{physics}
\usepackage{graphicx}
\usepackage{booktabs}
\usepackage[caption=false]{subfig}

\usepackage{epsfig}
\usepackage{listings}
\usepackage{multirow}
\usepackage{colortbl}

\newcommand{\pdf}[1]{{\text{PDF}#1}}

\makeatletter

    \def\CT@@do@color{%
      \global\let\CT@do@color\relax
            \@tempdima\wd\z@
            \advance\@tempdima\@tempdimb
            \advance\@tempdima\@tempdimc
    \advance\@tempdimb\tabcolsep
    \advance\@tempdimc\tabcolsep
    \advance\@tempdima2\tabcolsep
            \kern-\@tempdimb
            \leaders\vrule
                    \hskip\@tempdima\@plus  1fill
            \kern-\@tempdimc
            \hskip-\wd\z@ \@plus -1fill }
    \makeatother

 \DeclareFontFamily{U}{wncy}{}
    \DeclareFontShape{U}{wncy}{m}{n}{<->wncyr10}{}
    \DeclareSymbolFont{mcy}{U}{wncy}{m}{n}

\begin{document}

\title{The vacuum provides quantum advantage to otherwise simulatable architectures}

\author{Cameron Calcluth}
\email{calcluth@gmail.com}
\affiliation{Department of Microtechnology and Nanoscience (MC2), Chalmers University of Technology, SE-412 96 G\"{o}teborg, Sweden}
\author{Alessandro Ferraro}
\affiliation{Centre for Theoretical Atomic, Molecular and Optical Physics, Queen's University Belfast, Belfast BT7 1NN, United Kingdom}
\affiliation{Dipartimento di Fisica ``Aldo Pontremoli,''
Università degli Studi di Milano, I-20133 Milano, Italy
}
\author{Giulia Ferrini}
\affiliation{Department of Microtechnology and Nanoscience (MC2), Chalmers University of Technology, SE-412 96 G\"{o}teborg, Sweden}

\begin{abstract}
We consider a computational model composed of ideal Gottesman-Kitaev-Preskill stabilizer states, Gaussian operations ---including all rational symplectic operations and all real displacements ---, and homodyne measurement. We prove that such architecture is classically efficiently simulatable, by explicitly providing an algorithm to calculate the probability density function of the measurement outcomes of the computation. We also provide a method to sample when the circuits contain conditional operations. This result is based on an extension of the celebrated Gottesman-Knill theorem, via introducing proper stabilizer operators for the code at hand. We conclude that the resource enabling quantum advantage in the universal computational model considered by B.Q. Baragiola {\it et al.} [Phys. Rev. Lett. \textbf{123}, 200502 (2019)], composed of a subset of the elements given above augmented with a provision of vacuum states, is indeed the vacuum state. 
\end{abstract}

\maketitle

\section{Introduction}
Identifying the physical resources underlying quantum advantage --- i.e., yielding the ability of quantum computers to solve computational problems faster than classical computers --- is of crucial importance for the design of meaningful architectures for quantum computation (QC)~\cite{Gour-2019}. Often, the resource depends on the model. For example, for architectures over finite-dimensional systems, Clifford circuits are resourceless from a computational standpoint, since they are efficiently simulatable~\cite{gottesman1997, gottesman1999,nielsen2000} until a so-called magic resource is provided, such as the T-state, which allows universal quantum computation to be performed~\cite{bravyi2005, reichardt2005}. Similarly, for infinite-dimensional continuous-variable (CV) systems, Gaussian circuits are efficiently simulatable~\cite{bartlett2002,mari2012,veitch2012} and to promote them to universal QC specific non-Gaussian resources~\cite{albarelli2018,takagi2018} have to be provided, such as the cubic-phase state~\cite{gottesman2001, lloyd1999}, or Gottesman-Kitaev-Preskill (GKP) states~\cite{baragiola2019,yamasaki2020}. The cost of producing these enabling resources with sufficient quality generally requires a significant overhead and their distinct features are typically complex and in stark contrast with respect to the elements of the corresponding simulatable architectures. For example, T-states and cubic-phase states are non-stabilizer and non-Gaussian, respectively. It is a natural question to ask: are   resources  always complex and costly to produce?

In this work, we provide a specific example of a CV quantum computing architecture that is classically efficiently simulatable, and that becomes universal by adding the vacuum state. The latter state is widely regarded as the simplest quantum state of a bosonic field, and in particular it is a Gaussian state. The architecture considered is based on stabilizer GKP states, Gaussian operations including conditional displacements and homodyne detection. By taking inspiration from stabilizer methods developed for discrete-variable (DV) systems~\cite{gottesman1997, gottesman1999,nielsen2000,beaudrap2013,gheorghiu2014}, we prove that this class of circuits is classically efficiently simulatable for rational symplectic operations and arbitrary continuous displacement, thereby significantly extending~\footnote{Note that in the main text, we simplify the class of simulatable operations to those which have a rational symplectic matrix. However, the class of simulatable operations also includes those given in the multimode case of Ref.~\cite{calcluth2022}. We provide the broader requirements of the class of simulatable symplectic matrices in Appendix~\ref{appendix:rational-projector-irrational-symplectic}}
the class of Gaussian operations that was previously known to be simulatable in combination with GKP states~\cite{garcia-alvarez2020,calcluth2022}. This result is obtained despite the fact that GKP states are highly non-Gaussian and their Wigner function is highly  negative~\cite{gottesman2001, garcia-alvarez2019, yamasaki2020}, and hence the standard theorems based on Gaussianity~\cite{bartlett2002} or  on the positivity of quasi-probability distributions~\cite{mari2012,veitch2012, rahimi-keshari2016}  cannot be applied. We then leverage on the results of Ref.~\cite{baragiola2019}, where the same architecture combined with the vacuum (or a thermal) state was shown to be universal for quantum computation, to conclude that the vacuum provides quantum advantage. 

The paper is structured as follows. In Sec.~\ref{sec:background} we provide an introduction to the circuit class that we demonstrate to be efficiently simulatable. In Sec.~\ref{sec:method-pdf} we provide an analytic method to evaluate the PDF of the introduced circuit class. Then, in Sec.~\ref{sec:efficient-algorithm} we provide an algorithm to evaluate the PDF of the circuit and show that it is classically efficient. We also extend our result to include adaptive circuits and show that GKP-encoded Clifford circuits are included in the simulatable class. We then demonstrate, in Sec.~\ref{sec:vacuum}, that these results are sufficient to conclude that the vacuum is a resource for quantum advantage in the context of the simulatable model we consider. In Sec.~\ref{sec:realistic-gkp-states-resource} we also extend this result to show that realistic GKP states can be considered resourceful in the context of this model. Finally, we provide conclusions and open questions in Sec.~\ref{sec:conclusion}.

\section{Gaussian circuits with stabilizer GKP states}
\label{sec:background}
In this section we introduce the circuit class considered in this work, which we later show to be efficiently simulatable.
\begin{figure}[h!]
     \centering
     \includegraphics[width=0.9\linewidth]{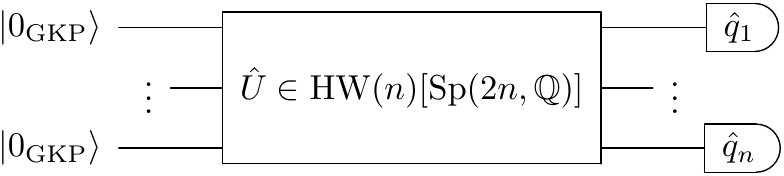}
     \caption{Schematics of the circuit class considered. In input, there are ideal GKP stabilizer states, such as the 0-logical state. The operations considered are the semi-direct product of the rational symplectic operations and the Heisenberg-Weyl group. Multimode homodyne detection follows. }
     \label{fig:circuitclass-maintext}
    \end{figure}
We consider the circuits shown in Fig.~\ref{fig:circuitclass-maintext}, where the input states are $n$ ideal GKP states encoding pure stabilizer states. Without loss of generality, we can consider each mode to be in the 0-logical encoded GKP state, which has a wave function in position representation given by~\cite{gottesman2001}
    \begin{align}
        \psi_{0,L}(x)=\bra{\hat q=x}\ket{0_{\text{GKP}}}=\sum_m \delta(2m\sqrt\pi - x);
    \end{align}
    the total multimode input state can be compactly indicated by 
    \begin{equation}
    \label{input-state}
        \ket{\mathbf 0_{\text{GKP}}}=\ket{0_{\text{GKP}}}^{\otimes n}.
    \end{equation} 
    The input state is stabilized by any combination of the operators $e^{2i\sqrt\pi \hat p_j},e^{i\sqrt\pi \hat q_j}$ with any integer power. This means that the action of these operators, or any combination of them, on the state will have the effect of the identity, e.g.
    \begin{align}
        e^{2i\sqrt\pi \hat p_j}\ket{\mathbf 0_{\text{GKP}}}=&\ket{\mathbf 0_{\text{GKP}}} \quad \forall j\in\{1,\dots,n\}\\
        e^{i\sqrt\pi \hat q_j}\ket{\mathbf 0_{\text{GKP}}}=&\ket{\mathbf 0_{\text{GKP}}}\quad \forall j\in\{1,\dots,n\}.
    \end{align}

The operations we consider in this work are those which belong to the group $\text{HW}(n)[\text{Sp}(2n,\mathbb Q)]$ which is the semi-direct product~\footnote{The Heisenberg-Weyl group $\text{HW}(n)$ is a normal subgroup of
the semi-direct product of $\text{HW}(n)$ and $\text{Sp}(2n,\mathbb Q)$, which we indicate by $\text{HW}(n)[\text{Sp}(2n,\mathbb Q)]$. Indeed, the subgroup $\text{HW}(n)$ is invariant under conjugation by any element of $\text{HW}(n)[\text{Sp}(2n,\mathbb Q)]$. Therefore, the full group of simulatable operations is specified by the semi-direct product of these two subgroups~\cite{dummit1991}.} of the Heisenberg-Weyl group $\text{HW}(n)$ and the rational symplectic group $\text{Sp}(2n,\mathbb Q)$. The Heisenberg-Weyl group $\text{HW}(n)$ consists of all real phase-space displacements of the form $e^{ic_j\hat q_j}$ and $e^{-id_j \hat p_j}$ for $c_j,d_j\in \mathbb R$ and $j\in\{1,\dots,n\}$.

The rational symplectic group  $\text{Sp}(2n,\mathbb Q)$ is the rational subgroup of the symplectic group $\text{Sp}(2n,\mathbb R)$ over the reals. It consists of all symplectic operations parameterized by a $2n\times 2n$ symplectic matrix such that all its elements are rational numbers. The set of rational symplectic operations is dense in the set of real symplectic operations. We provide proof of this fact in Appendix~\ref{appendix:density-proof}.  Note, however, that the density of the rational symplectic matrices should be regarded as a mathematical property characterizing the extent of the class of simulatable operations. It does not imply that the probability distributions obtained with operations parameterized by operations that are outside the set (e.g., in its closure) are necessarily simulatable.
For later convenience, we will denote a symplectic matrix $M$ by square sub-blocks of equal dimension:
\begin{align}
    \label{eq:block-def}
    M=\mqty(A&B\\C&D).
\end{align}

Gaussian operations can always be expressed as a unitary operator $\hat U$ in terms of symplectic operations and phase-space displacements~\cite{ferraro2005,serafini2017}.
The following operations form a generating set of all Gaussian operations:
\begin{equation}
\label{eq:generator-Gaussian}
    \{e^{ic_j\hat q_j},e^{i \theta_j(\hat q_j^2 + \hat p_j^2)/2},e^{-i\ln{s_j}(\hat q_j\hat p_j+\hat p_j\hat q_j)/2},e^{-i\hat q_j\hat p_k}\}
\end{equation}
where $c_j\in\mathbb R$, $\theta_j\in [0,2\pi),s_j\in\mathbb R$ and $j,k\in \{1,\dots,n\}$. These generators and also any combination of them will be shown to be simulatable so long as $\theta_j$ and $s_j$ are chosen such that $\cos\theta_j,\sin\theta_j,s_j\in\mathbb Q$ for all $j$. We will also show that adaptivity can be included as a feature of the class of circuits that can be efficiently simulated.

The circuits we consider are measured using homodyne detection, which, without loss of generality, we can restrict to position measurements. The measurement outcomes of the circuit in Fig.~\ref{fig:circuitclass-maintext} will therefore have a probability density function (PDF) expressed as
\begin{align}
    \label{eq:schrodinger-pdf}
    \pdf(\hat{\mathbf{q}}=\mathbf x)=\abs{\bra{\hat{\mathbf q}=\mathbf x}\hat U\ket{\mathbf 0_{\text{GKP}}}}^2.
\end{align}
When measuring the output modes, a quantum computer will provide outputs $\mathbf x$ selected with probabilities specified by the PDF in Eq.~(\ref{eq:schrodinger-pdf}). 

As we will clarify in Sec.~\ref{sec:vacuum}, the circuit elements (including adaptive operations) composing the universal model stemming from Ref.~\cite{baragiola2019} all belong to our class of circuits except for the vacuum.

\section{Simulation method for GKP circuits}
\label{sec:method-pdf}
In order to assess the simulatability of the circuits outlined in the previous section, we introduce a novel method to evaluate the PDF of the circuit presented in Fig.~{\ref{fig:circuitclass-maintext}}. This method involves tracking the Heisenberg evolution of the  measurement operators and then using the stabilizers of the input states to evaluate the PDF. We first provide an overview of the problem statement and a summary of the contents of the following subsections, which contain details of the proof.

A general Gaussian operation $\hat U$ belonging to $\text{HW}(n)[\text{Sp}(2n,\mathbb Q)]$ transforms, in the Heisenberg picture, the measurement operators $\hat q_j$ according to~\cite{bartlett2002,kok2010}
\begin{align}
\label{eq:evolution-operators-q}
    \hat Q_j=\hat U^\dagger \hat q_j \hat U=\sum_i a_{i}^{(j)}\hat q_i+b_{i}^{(j)}\hat p_i+c_j
\end{align}
where the coefficients $a_{i}^{(j)}=A_{i,j}$ and $b_{i}^{(j)}=B_{i,j}$ are elements of the blocks of the symplectic matrix $M$ as defined in Eq.~(\ref{eq:block-def}).  The vector $\vec c\in \mathbb R^n$, with elements $c_j$, describes the displacement in position. As we now prove, these circuits can be simulated in the strong sense by calculating the PDF. The PDF given in Eq.~(\ref{eq:schrodinger-pdf}) can be written in the Heisenberg picture using Eq.~(\ref{eq:evolution-operators-q}) as
\begin{align}
    \label{eq:fullpdf-maintext}
    \pdf(\hat{\mathbf Q}=\mathbf x)=\bra{\mathbf 0_{\text{GKP}}}\left(\prod_j \ket{\hat Q_j=x_j}\bra{\hat Q_j=x_j}\right) \ket{\mathbf 0_{\text{GKP}}}.
\end{align}

Our method is based on two main observations. First, by inserting the GKP stabilizers $e^{2i\sqrt\pi \hat p_j}$ and $e^{i\sqrt\pi \hat q_j}$ into the expression Eq.~(\ref{eq:fullpdf-maintext}), we can identify a periodicity relation of the PDF. Second, we can manufacture bespoke additional stabilizers, in terms of the Heisenberg measurement operators, of the form
\begin{align}
\label{eq-stabiliser-g}
    g(\vec l)=&e^{i\phi(\vec l)}\prod_j e^{i\sqrt\pi l_j \hat Q_j}
\end{align}
where $\vec l$ is an $n$-vector of real coefficients $l_j$ and $\phi(\vec l)$ is a phase factor chosen such that $g(\vec l)$ is a stabilizer. By inserting this bespoke stabilizer into the PDF, it is possible to identify a second constraint that provides the non-zero values of the PDF. Together, these two constraints uniquely identify the PDF.

In Sec.~\ref{sec:periodicity} we demonstrate how to derive the periodicity condition on the PDF, from the symplectic matrix $M$. In Sec.~\ref{sec:non-zero} we demonstrate how to identify the non-zero points of the PDF.
Finally, in Sec.~\ref{sec:non-zero-equal} we demonstrate that these two conditions are sufficient to construct the PDF of the circuit. A reader uninterested in the technical derivations may proceed directly to Eq.~(\ref{eq:final-pdf}), whereby we provide the explicit PDF of the circuit shown in Fig.~\ref{fig:circuitclass-maintext}. This PDF will provide sufficient information to understand the next Section \ref{sec:efficient-algorithm}, whereby we provide the algorithm to simulate these circuits.

\subsection{Periodicity of the PDF}
\label{sec:periodicity}
In this subsection, we will evaluate a periodicity condition that will provide a restriction on the PDF of the circuits considered. This periodicity condition informs us of the points of the PDF for which the values of the PDF are equal. The PDF, as given in Eq.~(\ref{eq:fullpdf-maintext}), can equivalently be written as
\begin{align}
    \label{eq:fullpdf-delta}
    \pdf(\hat{\mathbf Q}=\mathbf x)
    =&\bra{\mathbf 0_{\text{GKP}}}\left(\prod_j\delta(\hat Q_j-x_j)\right) \ket{\mathbf 0_{\text{GKP}}},
\end{align}
whereby we have rewritten the measurement operator as a delta function, i.e.
\begin{align}
    \label{eq:measurement-proj-delta}
     \delta(\hat Q_j-x_j)=&
    \int \dd s e^{is(\hat Q_j-x_j)}.
\end{align}
Similarly to the original Gottesman-Knill theorem for qubits~\cite{gottesman1997,nielsen2000}, by inserting stabilizers into this PDF on the right-hand-side of the delta function, and then using commutation relations to move the stabilizers to the left-hand-side, we find two expressions for the PDF which are equivalent. These two expressions correspond to two separate points on the PDF, implying that the PDF is equal at these points.
We start by considering the commutation of a general stabilizer with the measurement projection operators. We would like to calculate how the stabilizers $e^{2i\sqrt\pi \hat p_k}$ and $e^{i\sqrt\pi \hat q_k}$ commute with the general measurement projector, given in Eq.~(\ref{eq:measurement-proj-delta}).

This can be calculated by using the Baker-Campbell-Hausdorff (BCH) formula~\cite{sakurai,gerry2005} for linear combinations of quadrature operators
\begin{align}
    &e^{\hat X+\hat Y+\frac 1 2 [\hat X,\hat Y]}=e^{\hat X}e^{\hat Y}\label{eq:bch}\\
    \implies & e^{\hat X}e^{\hat Y}=e^{\hat Y}e^{\hat X}e^{[\hat X,\hat Y]} \label{eq:bchbraid},
\end{align}
valid for the case in which the operators ${\hat X}$ and ${\hat Y}$ commute with their commutator. The commutation between the measurement projector in Eq.~(\ref{eq:measurement-proj-delta}) and each stabilizer can be evaluated using Eq.~(\ref{eq:bchbraid}) by first evaluating how the terms commute, without integration. For the stabilizer containing $\hat p_k$ we find
\begin{align}
    e^{is(\hat Q_j-x)} e^{2i\sqrt\pi \hat p_k}=&e^{2i\sqrt\pi \hat p_k}e^{is(\hat Q_j-x)}e^{[is(\hat Q_j-x),2i\sqrt\pi \hat p_k]}\nonumber\\
    =&e^{-2s\sqrt\pi a^{(j)}_k i}e^{2i\sqrt\pi \hat p_k}e^{is(\hat Q_j-x)}    \label{eq:commutationmom},
\end{align}
whereas, for the stabilizer containing $\hat q_k$, we find
\begin{align}
    e^{is(\hat Q_j-x)} e^{i\sqrt\pi \hat q_k}=&e^{i\sqrt\pi \hat q_k}e^{is(\hat Q_j-x)}e^{[is(\hat Q_j-x),i\sqrt\pi \hat q_k]}\nonumber\\
    =&e^{s\sqrt\pi b^{(j)}_k i}e^{i\sqrt\pi \hat q_k}e^{is(\hat Q_j-x)}
    \label{eq:commutationpos}.
\end{align}
The first of these relations, Eq.~(\ref{eq:commutationmom}), allows us to calculate the commutation between the measurement projection operator and any integer $m_k\in\mathbb Z$ power of the momentum stabilizer $e^{2im_k\sqrt\pi \hat p_k}$, i.e.,
\begin{align}
    &\delta(\hat Q_j-x_j)e^{2im_k\sqrt\pi \hat p_k}\nonumber \\
    =&\int \dd s e^{-2m_ks\sqrt \pi a_{k}^{(j)}  i} e^{2im_k\sqrt\pi \hat p_k}e^{is(\hat Q_j-x_j)}\nonumber\\
    =&e^{2im_k\sqrt\pi \hat p_k}\delta(\hat Q_j-x_j-2m_k\sqrt\pi a_{k}^{(j)}).
\end{align}
The second relation  Eq.~(\ref{eq:commutationpos}) provides us with a similar relation for any integer $m_k'\in\mathbb Z$ power of the position stabilizer $e^{im_k'\sqrt\pi \hat q_k}$, i.e.,
\begin{align}
    \delta(\hat Q_j-x_j)e^{-im_k'\sqrt\pi \hat q_k}=&e^{-im_k'\sqrt\pi \hat q_k}  \delta(\hat Q_j-x_j-m_k'\sqrt\pi b_{k}^{(j)}).
\end{align}
Now, inserting all the stabilizers
\begin{align}
    e^{2\sqrt\pi i m_1\hat p_1}e^{-\sqrt\pi i m_1'\hat q_1}\dots e^{2\sqrt\pi i m_n\hat p_n}e^{-\sqrt\pi i m_n'\hat q_n}
\end{align}
to the right-hand side of full the PDF given by Eq.~(\ref{eq:fullpdf-delta}), and using the commutation relations to move the stabilizers to the left-hand side, we find that the PDF at $\vec x$ is equal to the PDF at the displaced point $\hat x'$, which can be expressed as
\begin{align}
    \label{eq:periodicitymultimode}
    x_j\to x_j'=x_j+\sqrt\pi \sum_k 2a_{k}^{(j)}m_k+b_{k}^{(j)}m'_k.
\end{align}

We also note that this periodicity condition can equivalently be written in terms of
\begin{align}
    \vec x' =&\vec x+ 2 \sqrt\pi \mqty(A&\frac 1 2 B)\mqty(\vec m\\ \vec m')
    \label{eq:appendix-periodicity}
\end{align}
where $\vec m$ and $\vec m'$ are each $n$-dimensional vectors of integers. This form of the periodicity relation will be useful when combining the two conditions in Sec.~\ref{sec:non-zero-equal}. This provides us with the first condition for the form of the PDF.
In the following subsection, we derive the second condition, which informs us of the set of points at which the PDF is non-zero.

\subsection{Set of non-zero points}
\label{sec:non-zero}
To evaluate the non-zero points of the PDF, we construct bespoke stabilizers from the Heisenberg measurement operators. Although this step varies notably from the conventional Gottesman-Knill theorem for DV systems, we can adopt a comparable approach to the DV theorem once such stabilizers have been identified. Specifically, we insert the stabilizers into the PDF and derive a set of equalities with respect to a set of points $\vec x$. These equalities lead to a contradiction unless the value of the PDF is only non-zero at this set of points.

We begin by identifying stabilizers of the set of input $0$-logical states, $\ket{\mathbf 0_{\text{GKP}}}$, expressed in terms of the Heisenberg measurement operators $\hat Q_j$. To do so, we first define an operator $g(\vec l)$, which will become a stabilizer for the input $0$-logical states under certain conditions. Considering a generic vector $\vec l\in \mathbb Q^n$ and a generic real function $\phi(\vec l): \mathbb Q^n \to \mathbb R$, the operator is defined as
\begin{align}
    g(\vec l)=&e^{i\phi(\vec l)}\prod_j e^{i\sqrt\pi l_j \hat Q_j}\nonumber\\
    =&e^{i\phi(\vec l)}  e^{i\sqrt\pi \sum_j l_j\left( \sum_k \left(A_{j,k}\hat q_k+B_{j,k}\hat p_k\right)+c_j\right)}\nonumber\\
    =&e^{i\phi(\vec l)} e^{i\sqrt\pi \vec l\cdot \vec c} \prod_k e^{i\sqrt\pi  \left(\sum_j l_jA_{j,k}\right)\hat q_k+i\sqrt\pi \left(\sum_jl_jB_{j,k}\right)\hat p_k}\nonumber\\
    =&e^{i\phi(\vec l)} e^{i\sqrt\pi \vec l\cdot \vec c} \prod_{k=1}^n e^{i\sqrt\pi  \left(\vec l^T A \right)_{k}\hat q_k+i\sqrt\pi \left(\vec l^T B\right)_{k}\hat p_k}.
\end{align}
Using the BCH formula given in Eq.~(\ref{eq:bch}) we find that each term in the product can be expressed as
\begin{align}
    e^{i\sqrt\pi  \left(\vec l^T A\right)_{k}\hat q_k}e^{i\sqrt\pi \left(\vec l^T B\right)_{k}\hat p_k}e^{\frac i 2\pi \left(\vec l^T A\right)_{k}\left(\vec l^T B\right)_{k} }.
\end{align}
and we can therefore express the operator as
\begin{align}
    g(\vec l)=&e^{i\phi(\vec l)} e^{i\sqrt\pi \vec l\cdot \vec c}e^{\frac i 2\pi \vec l^T AB^T\vec l} \prod_{k=1}^n e^{i\sqrt\pi  \left(\vec l^T A\right)_{k}\hat q_k}e^{i\sqrt\pi \left(\vec l^T B\right)_{k}\hat p_k}.
\end{align}
We find that by choosing $\phi(\vec l)$ to be
\begin{align}
\label{eq:phi-form}
    \phi(\vec l)
    =&-\frac 1 2\pi \vec l^T AB^T\vec l-\sqrt\pi \vec l\cdot \vec c,
\end{align}
this operator will have the form
\begin{align}
    g(\vec l)
    =&
    \prod_k e^{i\sqrt\pi  \left(\vec l^T A\right)_{k}\hat q_k}e^{i\sqrt\pi \left(\vec l^T B\right)_{k}\hat p_k}.
\end{align}
Hence, $g(\vec l)$ will be a stabilizer of $\ket{\mathbf 0_{\text{GKP}}}$ whenever
\begin{align}
    \label{eq:condition1}
    (A^T\vec l)_{k}=&0\mod 1\nonumber \\
    (B^T\vec l)_{k}=&0\mod 2.
\end{align}

Inserting the stabilizer $g(\vec l)$ into the equation of the PDF, given in Eq.~(\ref{eq:fullpdf-delta}), we have an equality between the PDF in its original form, and the PDF with the inserted stabilizer. Specifically, by inserting the stabilizer between the Heisenberg-evolved position quadrature basis states and the $0$-logical GKP states, we find that the stabilizer will act on the basis states as
\begin{align}
     &g(\vec l)\prod_j\ket{\hat Q_j=x_j}\bra{\hat Q_j=x_j}\nonumber\\
     =&e^{i\phi(\vec l)}\prod_j e^{i\sqrt\pi l_j x_j}\ket{\hat Q_j=x_j}\bra{\hat Q_j=x_j},
\end{align}
where the choice of $\vec l$ is constrained by Eq.~(\ref{eq:condition1}) and $\phi(\vec l)$ is of the form given in Eq.~(\ref{eq:phi-form}).
Furthermore, given that we know that the PDF will be equal, with or without the inserted stabilizer, we find that
\begin{align}
     &\bra{\mathbf 0_{\text{GKP}}}\left(\prod_j\ket{\hat Q_j=x_j}\bra{\hat Q_j=x_j}\right)\ket{\mathbf 0_{\text{GKP}}}\nonumber\\
     =&e^{i\phi(\vec l)}\prod_j e^{i\sqrt\pi l_j x_j}\bra{\mathbf 0_{\text{GKP}}}\ket{\hat Q_j=x_j}\bra{\hat Q_j=x_j}\ket{\mathbf 0_{\text{GKP}}}.
\end{align}

This equality can only be true if the term involving the phase equals $1$, or the PDF itself is zero. Hence, the non-zero points $\vec x$ of the PDF satisfy the equation
\begin{align}
    \label{eq:condition2}
    \sqrt\pi \vec l^T \vec x-\frac 1 2 \pi \vec l^T AB^T\vec l-\sqrt\pi \vec l\cdot \vec c=0\mod 2\pi
\end{align}
for all possible choices of $\vec l$ which satisfies Eq.~(\ref{eq:condition1}). 
If, on the other hand, we choose a different point $\vec x$, that does not satisfy this constrained equation, the equality will result in a contradiction unless the PDF is zero at these values of $\vec x$. We can therefore reduce the problem of identifying the non-zero points to finding solutions $\vec x$ of the equation Eq.~(\ref{eq:condition2}) constrained by Eq.~(\ref{eq:condition1}). We now provide a short summary of the steps required to solve Eq.~(\ref{eq:condition2}) given Eq.~(\ref{eq:condition1}), provided that $A$ and $B$ both contain all rational elements. The full details are provided in Appendix~\ref{sec:appendix-solution-constrained-eq}.

To solve this constrained equation we first find the allowed vectors $\vec l$. This can be achieved by introducing the matrix $S$ which is defined as
\begin{align}
    \label{eq:def-S}
    S=\mqty(A^T \\ \frac 1 2 B^T).
\end{align}
Then, the constraint on the allowed values of $\vec l$ is given by $S\vec l=\vec b$ where $\vec b$ is a vector of $2n$ integers. The Moore-Penrose pseudoinverse $S^+$ provides a method to find solutions of the form $\vec l=S^+\vec b$~\cite{moore1920,penrose1955,ben2003}. The solutions of $\vec l$ can be found by first finding the Smith decomposition~\cite{newman1972,newman1997,ben2003} of the matrix $\sigma S$, where $\sigma$ is the smallest integer for which the elements of the matrix $\sigma S$ are all integers. Note that this step assumes that the symplectic matrix, and therefore also $S$, is rational. We provide broader requirements for the symplectic matrix in Appendix~\ref{appendix:rational-projector-irrational-symplectic} and discuss the relationships between these classes of simulatable operations in Appendix~\ref{appendix:relationships}. Next, using the Smith decomposition of $\sigma S=VDU$ we identify which integer choices of $\vec b$ will provide valid solutions of $\vec l$. We find that the vectors $\vec l$ can be expressed as~\cite{stackexchangesmith}
\begin{align}
    \label{eq:l-solution}
    \vec l=R\vec m
\end{align}
where $\vec m$ is any choice of an $n$-vector of integers and $R$ is an $n\times n$ invertible rational matrix, defined as
\begin{align}
    \label{eq:def-r}
    R=S^+V\mqty(\mathbbm 1\\0).
\end{align}
We can then rewrite Eq.~(\ref{eq:condition2}) as a system of linear equations of the form
\begin{align}
    \frac{1}{\sqrt\pi}R^T(\vec x-\vec c)=\vec t \mod 2
    \label{eq:condition2-system}
\end{align}
where $\vec t$ is the main diagonal of the matrix $T=\frac 1 2 R^TAB^TR$.
This form, i.e.,  Eq.~(\ref{eq:condition2-system}), allows us to evaluate the solution to the constrained equation as
\begin{align}
    \label{eq:non-zero}
    \vec x=\sqrt\pi R^{-T}(\vec t+2\vec m)+\vec c,
\end{align}
Therefore, provided that the symplectic matrix is rational, we have identified that the PDF is non-zero exclusively at these points.

Combined with the insight from the previous subsection, we have now identified both a periodicity relation and the set of non-zero points of the PDF. In the following subsection, we use both these results to demonstrate that the PDF assumes the same value at all the non-zero points, i.e., those identified in Eq.~(\ref{eq:non-zero}).

\subsection{Constructing the PDF of the circuit}
\label{sec:non-zero-equal}

We now show that the PDF is specified completely by the periodicity relation and the points at which the PDF is non-zero. For this to hold, two conditions are required. First, we need to ensure that any non-zero point displaced by the periodicity relations always results in another non-zero point. Second, we need to ensure that any non-zero point can be reached by another non-zero point using the periodicity relations. 

We begin with the first condition and show that for any valid solution $\vec x$ we also get a valid solution if it is displaced according to the periodicity constraint. Namely, we can check that any point specified by the periodicity constraint is included in the allowed points.

If we take a point specified by 
\begin{align}
    \label{eq:chosen-point}
    \frac{1}{\sqrt\pi}\vec x^{(1)}=(R^T)^{-1}(\vec t+2\vec m)  +\vec c
\end{align}
and displace it according to the periodicity relation provided in Eq.~(\ref{eq:appendix-periodicity}), the new point should also satisfy this constraint. Here, to distinguish the vectors of integers in Eq.~(\ref{eq:appendix-periodicity}) and Eq.~(\ref{eq:chosen-point}), we relabel the arbitrary choice of integers in Eq.~(\ref{eq:appendix-periodicity}), given as vectors $\vec m$ and $\vec m'$, as $\vec k$ and $\vec k'$, respectively. Given a displacement, specified by $\vec k$ and $\vec k'$, we find a new point
\begin{align}
     \frac{1}{\sqrt\pi}\vec x^{(2)}=(R^T)^{-1}(\vec t+2\vec m)+2\mqty(A&\frac 1 2 B)\mqty(\vec k\\\vec k') +\vec c.
\end{align}
For the first condition to hold, this new point must also be a non-zero point of the PDF, and should satisfy the system of linear equations defining the non-zero points, given in Eq.~(\ref{eq:condition2-system}). This can be checked by inserting $\vec x^{(2)}$ into the left-hand side of that equation,
\begin{align}
    \label{eq:insert-x2}
    &\frac{1}{\sqrt\pi}R^T(\vec x^{(2)}-\vec c) \nonumber\\
    =&R^T\left((R^T)^{-1}(\vec t+2\vec m)+2 \mqty(A&\frac 1 2 B)\mqty(\vec k\\\vec k')\right)\nonumber\\
     =&\vec t+2\vec m+2 R^T\left(A\vec k+\frac 1 2 B\vec k'\right)
\end{align}
which we expect to evaluate to $\vec t+2\vec m'$, where $\vec m'$ is a different $n$-vector of integers.
This can be shown by inspecting each element of the vector that is given as the third term of Eq.~(\ref{eq:insert-x2}). We label this vector as $\vec w$, i.e.,
\begin{align}
    \label{eq:appendix-w-vector}
    \vec w=&2R^T\left(A\vec k+\frac 1 2 B\vec k'\right).
\end{align}
The elements of this vector can be found by multiplying its transpose with the unit vector
\begin{align}
    w_i=& \vec w^T\vec e_i\nonumber\\
    =&2 \left(\vec k^TA^TR+\frac 1 2 \vec k'^TB^TR\right)\vec e_i.
\end{align}
We know from Eq.~(\ref{eq:l-solution}) that for any $n$-dimensional vector of integers $\vec k$ there exists an allowed value of $\vec l$ as
\begin{align}
    \vec l=R\vec k.
\end{align}
Choosing $\vec k$ to be the basis vector $\vec k=\vec e^{(i)}$, which is zero in all entries except at $i$, we can identify one choice of $\vec l$, parameterized by $i$, that corresponds to the element of the vector $\vec k$, chosen to be non-zero, as
\begin{align}
    \vec l^{(i)}=R\vec e_i.
\end{align}
We can then write the $i$-th element of the vector $\vec w$ in Eq.~(\ref{eq:appendix-w-vector}) as
\begin{align}
\label{eq:appendix-w-vector-index}
    w_i
    =&2\left(\vec k^TA^T\vec l^{(i)}+\frac 1 2 \vec k'^TB^T\vec l^{(i)}\right).
\end{align}
Furthermore for any allowed $\vec l$, including the choice $\vec l^{(i)}$ we have
\begin{align}
    (A^T\vec l)_i=0\mod 1\nonumber\\
    (B^T\vec l)_i=0\mod 2.
\end{align}
The term in brackets in Eq.~(\ref{eq:appendix-w-vector-index}) must be an integer, and so $w_i$ must be an even integer. This means that
\begin{align}
    \vec w=2 \tilde{\vec m}
\end{align}
for some $n$-dimensional vector of integers $\tilde{\vec m}$ and hence 
\begin{align}
    \frac{1}{\sqrt\pi}R^T\vec x^{(2)}
    =\vec t+2\vec m+\vec w=\vec t+2\vec m',
\end{align}
which is of the same form as Eq.~(\ref{eq:condition2-system}). This implies that any non-zero point displaced using the periodicity condition also satisfies the constrained equation specifying the non-zero points. We have therefore demonstrated that the first condition introduced in this subsection does indeed hold.

For the second condition, we need to demonstrate that any non-zero point can be reached using the periodicity relations.

This can be proven by specifying a center point as
\begin{align}
    \vec x^{(0)}=\sqrt\pi (R^{T})^{-1}\vec t
\end{align}
and demonstrating that it can be displaced to any other non-zero point of the form
\begin{align}
    \vec x^{(1)}=\sqrt\pi (R^{T})^{-1}(\vec t+2\vec m)
\end{align}
using only displacements of the form given by Eq.~(\ref{eq:appendix-periodicity}). This is equivalent to saying that for any choice of $\vec m$, there exists some $\vec k,\vec k'$ such that 
\begin{align}
    &\sqrt\pi(R^{T})^{-1}(\vec t+2\vec m)=\sqrt\pi(R^{T})^{-1}\vec t+ 2 \sqrt\pi S^T\mqty(\vec k\\ \vec k')\nonumber\\
    \implies &(R^{T})^{-1}\vec m=  S^T\mqty(\vec k\\ \vec k').
\end{align}
We can solve this equation using the pseudoinverse to find potential solutions of the form
\begin{align}
    \mqty(\vec k\\ \vec k')=  (S^T)^+(R^{T})^{-1}\vec m.
\end{align}
As with any pseudoinverse, we can check whether this solution is a valid solution by evaluating whether the original linear equation holds under the solution. I.e. we check
\begin{align}
    S^T(S^T)^+(R^{T})^{-1}\vec m=(S^+S)^T(R^{T})^{-1}\vec m=(R^{T})^{-1}\vec m
\end{align}
which means that this solution is one possible valid solution. Note there exists infinite more solutions but we do not need to find an expression for all of these. We have shown that no matter which non-zero point we are interested in, i.e. $\vec x^{(1)}$, there will be at least one way --- and, in fact infinite, ways --- to get to that point from the center point $\vec x^{(0)}$. This completes the proof of the second condition, introduced in this subsection.

We have therefore shown that both conditions hold, meaning that any non-zero point displaced by the periodicity relations results in another non-zero point and that any non-zero point can be reached from any other non-zero point. This implies that the value of the PDF is equal for all the non-zero points specified in Eq.~(\ref{eq:non-zero}).

This allows us to write the full 
and exact PDF of the multimode measurement, in terms of these allowed points, as
\begin{align}
    \label{eq:final-pdf}
    \pdf(\vec x)=\sum_{\vec m\in \mathbb Z^n}\delta(\vec x-\sqrt\pi R^{-T}(\vec t+2\vec m)-\vec c).
\end{align}
As we will show, this method to evaluate the PDF can be implemented with an efficient algorithm --- namely, an algorithm whose complexity increases at most polynomially with respect to the number of modes.
The algorithm for computing this PDF, along with its complexity analysis, is provided in Sec.~\ref{sec:efficient-algorithm}.

\section{Efficient algorithm for the simulation of GKP circuits}
\label{sec:efficient-algorithm}
In this section we provide an explicit algorithm to evaluate the PDF of the circuit shown in Fig.~\ref{fig:circuitclass-maintext} and derive some notable consequences of this result.

Efficient classical computation of the PDF of a quantum circuit is referred to as strong simulation. 
We begin with a presentation of the algorithm to efficiently simulate the circuits shown in Fig.~\ref{fig:circuitclass-maintext} in Sec.~\ref{sec:algorithm-and-complexity} in the strong sense. We, therefore, extend the simulatable class to all real displacements and all rational symplectic operations as opposed to a restricted set~\footnote{Alternative previous results~\cite{Juani-thesis,bermejo-vega2016} also exist for the simulation of CV circuits in the form of normalizer circuits. These results provide a numerical method to simulate non-adaptive normalizer circuits in the weak sense~\cite{jozsa2014}, i.e. it is possible to sample the output of a non-adaptive circuit. However, adaptivity is required for magic state distillation and so these results alone do not allow us to conclude that the vacuum is responsible for providing quantum advantage.}. Furthermore, the size of the set of simulatable operations does not depend on the number of modes measured, as was the case in Ref.~\cite{calcluth2022}.

The complementary notion of weak simulatability means instead that a classical computer can efficiently sample the outcomes of the circuit~\cite{jozsa2014}. Weak simulation is sufficient to conclude that a quantum circuit will not provide quantum advantage, as a quantum computer will, in any case, produce outcomes selected from the PDF. Following the argument of Ref.~\cite{jozsa2014}, and assuming the capability of sampling from the set of integers, we will demonstrate in Sec.~\ref{sec:adaptive} that by restricting to weak simulation, we can further extend the class of simulatable circuits shown in Fig.~\ref{fig:circuitclass-maintext}. This extended class includes adaptive circuits, whereby intermediate measurement outcomes can affect future operations.

We will later use these results to demonstrate that the routine introduced in Ref.~\cite{baragiola2019} --- whereby the vacuum and GKP states are used to perform universal quantum computation --- is efficiently simulatable when the vacuum is removed. This circuit is adaptive and contains GKP-encoded Clifford operations.

With this motivation, we demonstrate, in Sec.~\ref{sec:clifford}, that GKP-encoded Clifford operations are included in the set of simulatable operations that we present in this work.
As a consequence, we can also now simulate all encoded qubit stabilizer GKP states as input states, in the same sense as the Gottesman-Knill theorem~\cite{gottesman1997,gottesman1999,nielsen2000}. This was not possible using our previous method~\cite{calcluth2022}.
Together, these results provide us with all the tools required to demonstrate, in the later Sec.~\ref{sec:vacuum}, that the vacuum is indeed the resource for quantum advantage in circuits composed of input GKP stabilizer states followed by Gaussian operations and homodyne measurement.

Finally, to demonstrate the practical implementation of the algorithm, we provide an example of evaluating the PDF of a simple circuit in Sec.~\ref{sec:algorithm-example}.
\subsection{Algorithm to evaluate the PDF}
\label{sec:algorithm-and-complexity}
We now provide the algorithm
to calculate the PDF of a general circuit shown in Fig.~\ref{fig:circuitclass-maintext} by using the result of the previous section. 
We will also provide an analysis of the computational time required to evaluate the PDF.

To express the PDF in Eq.~(\ref{eq:final-pdf}) given the symplectic matrix $M$, given in block form as defined in Eq.~(\ref{eq:block-def}), and the vector of displacement $\vec c$, we need to evaluate $R^{-T}$ and $\vec t$.
The matrix $R^{-T}$ is given in terms of $S^T$ and $V$, where $V$ is the unimodular matrix arising from the Smith decomposition of $\sigma S$. The vector $\vec t$ can be evaluated from $V$.

First, we identify the matrix $S$, by simply writing it in terms of the block components $A,B$ as it is given in Eq.~(\ref{eq:def-S}).
To find the matrix $V$ we first need to calculate the lowest common multiple of all the denominators of the elements $S$. Formally we could write
\begin{align}
    \label{eq:appendix-sigma-formal}
    \sigma = \text{lcm}(\text{den}(S)).
\end{align}
where $\text{den}(\cdot)$ evaluates the denominator of all matrix elements and $\text{lcm}(\cdot)$ evaluates the lowest common multiple of all matrix elements.

Then we multiply the matrix $S$ by $\sigma$ to produce an integer matrix $\sigma S$. We can perform a Smith normal form decomposition on this matrix to identify the $2n\times 2n$ unimodular matrix $V$, the $2n\times n$ diagonal matrix $D$ and the $n\times n$ unimodular matrix $U$,
\begin{align}
    \label{eq:appendix-direct-snf}
    \sigma S=VDU.
\end{align}
We can discard the matrices $D,U$.

The transpose-inverse of $R$ can be directly evaluated as
\begin{align}
    R^{-T}=S^TV^{-T}\mqty(\mathbbm 1 \\0)=\mqty(A& \frac 1 2 B)V^{-T}\mqty(\mathbbm 1 \\0).
\end{align}

Furthermore, the matrix $T$ can be calculated from $V$ as
\begin{align}
    T=&V^{(11)T} V^{(21)}
\end{align}
and the vector $\vec t$ is simply the diagonal entries of $T$. The PDF is then given by Eq.~(\ref{eq:final-pdf}).

To summarize, this algorithm consists of the following steps

\begin{enumerate}
    \item Evaluate the matrix $S$ from $M$
    \item Identify the integer $\sigma$ from Eq.~(\ref{eq:appendix-sigma-formal}).
    \item Multiply every element of $S$ by $\sigma$
    \item Find the matrix $V$ from the Smith decomposition of $\sigma S$
    \item Find the inverse-transpose of $V$
    \item Evaluate $R^{-T}$ from $S^T,V^{-T}$
    \item Evaluate $\vec t$ from $V$
\end{enumerate}

We can assume that the $2n\times 2n$ symplectic matrix $M$ is stored as a matrix of numerators $M^{\text{num}}$ and a matrix of denominators $M^{\text{den}}$ such that $M=M^{\text{num}}\oslash M^{\text{den}}$, where $\oslash$ denotes element-wise division.

Step 1 consists of a truncation of the $2n\times 2n$ matrix $M$ followed by matrix multiplication of the denominator matrix $M^{\text{den}}$ which in the worst case requires $\mathcal O(n^3)$ operations~\cite{arora2009}.

Step 2 consists of finding the lowest common multiple of every element in $M^{\text{den}}$. There are $(2n)^2$ integer entries of this matrix $M^{\text{den}}_{i,j}$. We can find the lowest common multiple of two integers $\alpha,\beta$ by using the greatest common divisor
\begin{align}
    \text{lcm}(\alpha,\beta)=\frac{\alpha\beta}{\text{gcd}(\alpha,\beta)}
\end{align}
and then calculate the lowest common divisor of more than $2$ integers iteratively, i.e.,
\begin{align}
    \text{lcm}(\alpha,\beta,\gamma)=\text{lcm}(\text{lcm}(\alpha,\beta),\gamma).
\end{align}

If we limit the number of digits of precision in each element of $M^{\text{den}}_{i,j}$ to $k$, we can identify that the calculation of the lowest common multiple of two integers of bit length $k$ will require at most $\mathcal O(k^2)$ operations~\cite{mollin2008,arora2009}. The size of the bit string representing the lowest common multiple will be at most $2k$. Calculating the lowest common multiple of two numbers of size $k,2k$ has complexity in terms of the bit length of the smallest of the two numbers, $k$ and so the complexity of calculating the next iteration will also be $\mathcal O(k^2)$ and the resulting lowest common multiple of the three numbers will be $3k$. We need to repeat this iterative process $n^2$ times and so the total time complexity will be in the worst case $\mathcal O(n^2k^2)$ and the size of the integer $\sigma$ will have at most $n^2 k$ bits.

Step 3 consists of multiplying every element of $S$ by $\sigma$ which will require $\mathcal O(n^2)$ operations and the matrix $\sigma S$ will contain $2n^2$ elements each of maximum size $n^2k+k$. Therefore, the bit length of each element of $\sigma S$ is polynomial in the number of modes $n$ considered.

Step 4 consists of finding a Smith normal form decomposition which is polynomial in the size of the matrix $S$ and the number of bits of each element~\cite{storjohann2000}, which we know from Step 3 is also polynomial in the number of modes $n$. Therefore Step 4 can be computed in polynomial time.

The remaining steps consist of linear algebra operations (i.e. matrix inversion, matrix multiplication and matrix transposition) which are all known to be polynomial in the size of the matrices considered and the bit length of each element~\cite{arora2009}.

We can therefore conclude that the entire algorithm for evaluating the exact PDF of the circuit is polynomial in the number of modes $n$. This means that all rational symplectic operations and all continuous displacements in the circuits of the form in Fig.~\ref{fig:circuitclass-maintext} are strongly simulatable.

In the following subsection, we will demonstrate that our result can be extended to include adaptive circuits, when restricting to weak simulation.

\subsection{Adaptive circuits are weakly simulatable}
\label{sec:adaptive}
While in the previous subsection we demonstrated that the class of circuits shown in Fig.~\ref{fig:circuitclass-maintext} are strongly simulatable, we will now demonstrate that this class can be extended to adaptive circuits when restricting to weak simulation. Adaptive quantum circuits contain intermediate measurements that can then either be used as parameters in future operations or can be used in a classical subroutine to decide if or where Gaussian operations are applied.

Formally, we can express adaptive circuits as beginning with a unitary operation $\hat U_0$, acted on the input state, followed by a series of $K$ operations and measurements of the form ~\cite{jozsa2014}
\begin{align}
    \hat U_{j}(x_1,\dots,x_j)M_{i_j(x_1,\dots,x_{j-1})}(x_j),
\end{align}
where $j\in \{1,\dots,K\}$.
After applying the initial unitary operation $\hat U_0$, we measure the mode $i_1$ which gives the result $x_1$. Next, we act with the operator $\hat U_1(x_1)$ which is parameterized by the previous measurement result $x_1$. Following this, we measure mode $i_2(x_1)$. The mode which is measured, i.e. $i_2$, may also depend on the previous measurement result $x_1$. This continues up to an arbitrary number $K$ of sequences of operations and measurements.

We now demonstrate that it is possible to sample from the circuits we have shown to be simulatable, even when incorporating adaptivity, in polynomial time. By the same logic of Theorem 5 of Ref.~\cite{jozsa2014} we can consider each measurement as a single run of a reduced circuit. I.e., starting with the first measurement $M_{i_1}(x_1)$, where we measure the $i_1$-th mode, we simulate the Gaussian circuit $\hat U_1$ acting on the input states, followed by a measurement on the $i_1$-th mode. We know, from the previous subsection, that we can calculate the PDF of this circuit. Hence, we can also sample a random measurement outcome of this circuit.

Next, we simulate a new circuit consisting of the operation
\begin{align}
    \hat U_1(x_1)\hat U_0
\end{align}
using the measurement outcome of the previous simulation, to decide the Gaussian operation $\hat U(x_1)$. Measurement of $i_1$ and $i_2(x_1)$ will give a PDF of the form
\begin{align}
    \pdf(x_1,x_2)
\end{align}
for which we can input the simulated measurement outcome $x_1$ of the previous simulation, in order to get a PDF in terms of only $x_2$. Again, simulating a single measurement outcome of $x_2$ allows us to continue this procedure for the rest of the measurements of the circuit. Therefore the outcome of any adaptive Gaussian circuit, for which the non-adaptive circuits are strongly simulatable, is weakly simulatable.

As a complementary result --- albeit, not necessary to reach the conclusions of this paper --- we also show, in Appendix \ref{sec:appendix-adaptive-mod}, that it is also possible to efficiently simulate the outcomes of adaptive circuits with modulo homodyne measurement.

In order to prove the result in Sec.~\ref{sec:vacuum}, i.e., that the vacuum is a resource for quantum advantage, we must also show that adaptive circuits containing GKP-encoded Clifford operations are efficiently simulatable. In the following subsection, we demonstrate that this is indeed the case.

\subsection{Clifford circuits are contained in the rational symplectic operations}
\label{sec:clifford}
We now demonstrate that GKP-encoded Clifford circuits are contained within the set of operations that we have shown to be efficiently simulatable.
Qubit Clifford circuits consist of stabilizer qubit states, acted on by Clifford operations, followed by measurement in a stabilizer basis. Without loss of generality, we can consider these circuits to be initialized in $0$ eigenstates of the Pauli $\hat Z$ operator, followed by Clifford operations and measured in the $\hat Z$ basis. Encoding these circuits into the GKP formalism gives circuits which consist of states initialized as $0$-logical GKP states, acted on by encoded Clifford operations, followed by homodyne measurement in the position basis.

The Clifford operations acting over $n$ modes can be described in terms of the following set of generators
\begin{align}
    \left\{e^{i\hat q_j^2/2},\hat F_j,e^{-i \hat q_j\hat p_k} : j,k\in\{1,\dots,n\} \right\}
\end{align}
where the Fourier transform is defined as
\begin{align}
    \hat F_j=e^{i\pi(\hat q_j^2+\hat p_j^2)/4}.
\end{align}
Note that in the case of the qubit encoding, it is not necessary to introduce phase-space displacements, as the required displacements can be produced by combinations of the symplectic operations.

Inspecting the symplectic form of each of these operators provides a description of the symplectic matrices of all Clifford group operations. Analyzing the generators of single-mode Clifford group operations we have
\begin{align}
    \hat F: \mqty(0&-1\\1&0)\\
    e^{i\hat q_j^2/2}: \mqty(1&0\\1&1)\\
    e^{-i\hat q_j\hat p_j}: \mqty(1&0&0&0\\1&1&0&0\\0&0&1&-1\\0&0&0&1).
\end{align}
By considering any combination of these operations we will clearly obtain only integer matrices.

The set of qubit Clifford operations can therefore be described as at least a subset of integer symplectic operations. Integer symplectic operations are contained within the class of rational symplectic operations. Therefore, all encoded qubit Clifford circuits are simulatable by our method.

This concludes our analysis of the types of circuits which are simulatable with our method. I.e., adaptive circuits consisting of input stabilizer GKP states, rational symplectic operations --- including GKP-encoded Clifford operations ---, real displacements and homodyne measurement, are all simulatable.

In the case of non-adaptive circuits, efficient strong simulation can be performed, whereby the PDF is evaluated efficiently. In the following subsection, we will apply this result to demonstrate the strong simulation of a simple circuit.

\subsection{Simple example}
\label{sec:algorithm-example}
We present an example of calculating the PDF of a simple circuit. We have specifically chosen a circuit that contains a vector $\vec t$ not equal to zero. We consider the circuit
\begin{align}
    \hat U=C_XF_1P_1^2F_1
\end{align}
where $P$ is the phase gate. Note that $F_1P_1^2F_1=X_1$ which means we would expect the action of this operator on two encoded qubits states to be $C_XX_1\ket{0_{\text{GKP}}}\ket{0_{\text{GKP}}}=\ket{1_{\text{GKP}}}\ket{1_{\text{GKP}}}$.

We can calculate its effect on the position measurement modes $\hat q_1$ and $\hat q_2$ as
\begin{align}
    \hat Q_1=\hat U^\dagger \hat q_1\hat U= -\hat q_1+2\hat p_1
\end{align}
and
\begin{align}
    \hat Q_2=\hat U^\dagger \hat q_2\hat U= -\hat q_1+2\hat p_1+\hat q_2
\end{align}
from which we can inspect
\begin{align}
    A=\mqty(-1&0\\-1&1) \quad \quad 
    B=\mqty(2&0\\2&0).
\end{align}
We can explicitly write the matrix $S$ as
\begin{align}
S=\mqty(A^T\\\frac 1 2 B^T)=\mqty(-1&-1\\0&1\\1&1\\0&0).
\end{align}
We then find the lowest common denominator of all the fractions of $S$. However, in this case, $\sigma=1$ since we already have all integers.
We can then calculate the Smith decomposition of $\sigma S=S$ which is given by
\begin{align}
    S=VDU
\end{align}
with
\begin{align}
    V=\mqty(-1&-1&0&0\\0&1&0&0\\1&1&1&0\\0&0&0&1) \quad D=\mqty(1&0\\0&1\\0&0\\0&0) \quad U=\mqty(1&0\\0&1).
\end{align}
We can also calculate the pseudoinverse of $S$ as
\begin{align}
    S^+=\mqty(-\frac 1 2&-1&\frac 1 2 &0\\0&1&0&0),
\end{align}
from which we can calculate $R$ as
\begin{align}
    R=S^+V\mqty(\mathbbm 1\\0)=\mqty(1&0\\0&1)
\end{align}
and $R^{-T}$ as
\begin{align}
    R^{-T}=\mqty(1&0\\0&1).
\end{align}
Furthermore we can find $T$ as
\begin{align}
    T=&\frac 1 2 R^TAB^TR\\
    =&\frac 1 2 \mqty(1&0\\0&1)\mqty(-1&0\\-1&1)\mqty(2&2\\0&0)\mqty(1&0\\0&1)\\
    =& \mqty(-1&-1\\-1&-1)
\end{align}
which gives the vector $\vec t$ of the diagonal elements of $T$ as
\begin{align}
    \vec t=\mqty(-1&-1)^T.
\end{align}
This allows us to express the PDF, which is given by
\begin{align}
    \pdf(\vec x)=&\sum_{\vec m\in \mathbb Z^n}\delta(\vec x-\sqrt\pi (R^T)^{-1}(\vec t+2\vec m)).
\end{align}
The PDF can be expressed in terms of each vector element of $\vec x$ as
\begin{align}
    &\pdf(\vec x)\nonumber \\
    =&\sum_{m_1,m_2\in \mathbb Z}\delta\left(x_1+\sqrt\pi -2\sqrt\pi m_1\right)\delta\left(x_2+\sqrt\pi-2\sqrt\pi m_2\right).
\end{align}
This is equivalent to measuring $\ket{1_{\text{GKP}}}$ in both modes, as we would expect, given the encoded circuit.

The results from this section provide us with the tools required to conclude the main result given in the following section, i.e., that the vacuum is the resource for quantum advantage in the context of this otherwise simulatable model.

\section{Vacuum yields quantum advantage}
\label{sec:vacuum}
\begin{figure}[h!]
    \centering
    \includegraphics[width=0.9\linewidth]{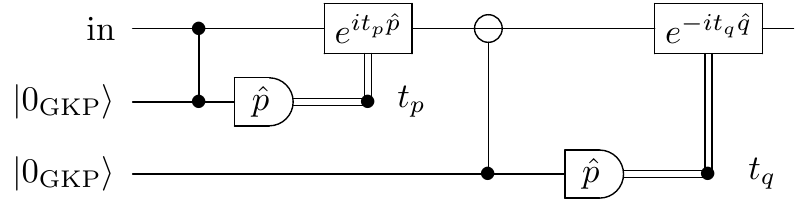}
     \caption{Circuit gadget implementing $\hat K_{\text{EC}}(\mathbf t)$. Mode $1$ is the top mode, which takes an input state and outputs a modified state. Mode $2$ and $3$ below are auxiliary modes which have a fixed input and once measured can be discarded. We use the notation of Ref.~\cite{noh2022low} whereby the controlled gate with the symbol $\ominus$ denotes the inverse of the SUM gate, namely $e^{i\hat q_3 \hat p_1}$. The measurement outcomes are denoted as $t_p$ and $t_q$.}
     \label{fig:kqgadget}
    \end{figure}

We now derive a notable consequence of the findings in the previous section, when combined with the results reported in Ref.~\cite{baragiola2019}. There, the circuit depicted in Fig.~\ref{fig:kqgadget} is used as the central resource to achieve magic-state distillation, and in turn fault-tolerant universality of an otherwise simulatable (GKP-encoded) stabilizer computation.
This circuit is composed of input GKP states $\ket{0_{\text{GKP}}}$, an additional CV input state (possibly the vacuum), GKP-encoded Clifford operations, homodyne measurements, displacements, and classical feed-forward of measurement results (Fig.~\ref{fig:kqgadget}). Such a circuit gadget has the effect to implement the Kraus operator ${\hat K_{\text{EC}}(\mathbf t) = \hat \Pi_{\text{GKP}}\hat V(-\mathbf t)}$, where ${\hat V(-\mathbf t)=e^{it_q\hat p}e^{-it_p\hat q}}$ and $\hat \Pi_{\text{GKP}}$ is the projection operator onto the GKP subspace. This has the effect of ``error correcting" the additional input state by projecting it onto the computational subspace of the GKP code. When the additional input state is the vacuum, this results in GKP-encoded magic states, except for a zero-measure set of the measurement outcomes $t_q$, $t_p$.

Performing this gadget across multiple modes with multiple additional vacuum states will provide a number of different states which each have a high fidelity to a magic $H$-type state. 

This gadget is adaptive and, once the auxiliary modes are measured, they can be discarded.
Within the gadget, the measurement values are used to shift the input state in position and momentum. Furthermore, the measurement outcomes give an indication of which $H$-type state the output state is closest to. We can use these measurement results to decide a Gaussian operation which shifts the state close to the target $\ket{\text{H}_{\text{GKP}}}$ state.

If we have $k$ copies of this gadget we will have produced $k$ different states which each has a high fidelity to the $\ket{\text{H}_{\text{GKP}}}$ state. We can then apply the twirling operation to each state, which for qubits is a probabilistic Clifford operation (hence implementable by a probabilistic Gaussian operation), which projects each state onto the $H$-axis of the Bloch sphere. These $k$ states are non-identical so require adaptive depolarizing operations, which are again probabilistic Clifford operations, to make all these $k$ states identical~\cite{privateben}. These adaptive probabilistic Clifford operations will adjust each state to match the state with the lowest fidelity to the target $H$-state. These operations are adaptive since they require knowledge of each state, which can be constructed from the values of $t_q,t_p$ measured for each gadget.

Now we have $k$ identical copies of states which have fidelity above the threshold for magic state distillation. The magic state distillation algorithm~\cite{bravyi2005,reichardt2005} involves Clifford operations, adaptive Clifford operations and probabilistic Clifford operations.
Therefore, the total algorithm to produce a $H$-type state from the vacuum and GKP states requires the following resources: input GKP states, input vacuum states, adaptive Clifford operations, probabilistic Clifford operations and homodyne measurements.

Now, consider the same procedure where instead of the vacuum state as the additional input, we only have $0$-logical (or another stabilizer) GKP states.

We know from this work that circuits involving GKP states, adaptive Clifford operations, probabilistic Clifford operations and homodyne measurements are weakly simulatable. Hence, the concatenation of all these operations, including the gadget in Fig.~\ref{fig:kqgadget}, belongs to the class of circuits that we have shown to be classically efficiently simulatable, if the supply of initial vacua is not included at the input of the circuit.
Therefore, in the context of the model of Ref.~\cite{baragiola2019}, ideal stabilizer GKP states, homodyne measurement, displacements and classical feed-forward of measurement outcomes are to be regarded as free operations, in the sense that they provide a simulatable model.

However, if we add the vacuum to this otherwise simulatable model, we find that it is promoted to universal quantum computation. We can thus conclude that the vacuum can be considered a resource for quantum advantage in this model.
Note that this conclusion was not possible to draw from Ref.~\cite{baragiola2019} solely, because the model considered, even excluding the additional vacuum state, was not proven to be classically efficiently simulatable therein. 

The intuition behind this result is that, as already noticed in Ref.~\cite{baragiola2019}, the interaction with the vacuum through an entangling operation takes the GKP states outside of the computational subspace spanned by the GKP logical codewords. Measurements followed by feed-forward and displacement project the unmeasured system back onto the GKP-encoded computational subspace, now in a magic state (apart from measurement outcomes which represent a zero-measure set in the set of all possible real measurement outcomes).

In the following section, we will extend this argument to demonstrate that realistic GKP states can also be considered a resource for quantum advantage in the context of this model.

\section{Realistic GKP states are a resource for quantum advantage}
\label{sec:realistic-gkp-states-resource}
Following the result of the previous section, we now demonstrate that realistic GKP states can also be considered a resource for quantum advantage.
Using a realistic (i.e., finitely squeezed) GKP state as the additional input state of the gadget in Fig.~\ref{fig:circuitclass-maintext}, instead of the vacuum, also produces a magic state with fixed probability, dependent on the squeezing of the realistic GKP state.

\begin{figure}[ht]
        \includegraphics[width=0.9\linewidth]{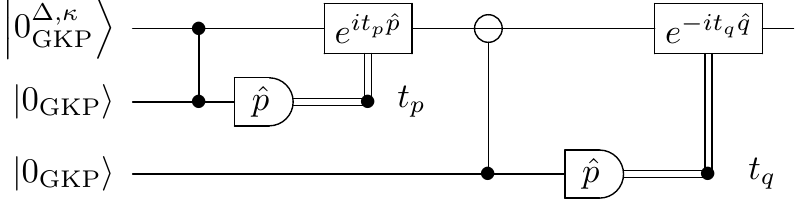}
         \caption{The error correcting circuit from Fig.~\ref{fig:kqgadget} implementing $\hat K_{\text{EC}}(\mathbf t)$, acting on an additional input non-ideal GKP state parameterized by $\Delta,\kappa$. Note that the control gate with the symbol $\ominus$ at the target is the inverse of the SUM gate, namely $e^{i\hat q_3 \hat p_1}$ \cite{noh2022low}.}
         \label{fig:nonidealgkpinput}
    \end{figure}

We now explicitly compute the outcome of the circuit in Fig.~\ref{fig:nonidealgkpinput}. Here, the additional input state, to be combined with ideal GKP states, is not the vacuum state, but instead a GKP state with variable squeezing. Note that the case of vacuum is re-obtained with a very good approximation by taking the limit of no squeezing in the GKP state. We will then compute the fidelity of the output state with the closest magic $\ket{H}$-type state.

    \begin{figure*}[t]
        \centering
        \includegraphics[width=2\columnwidth]{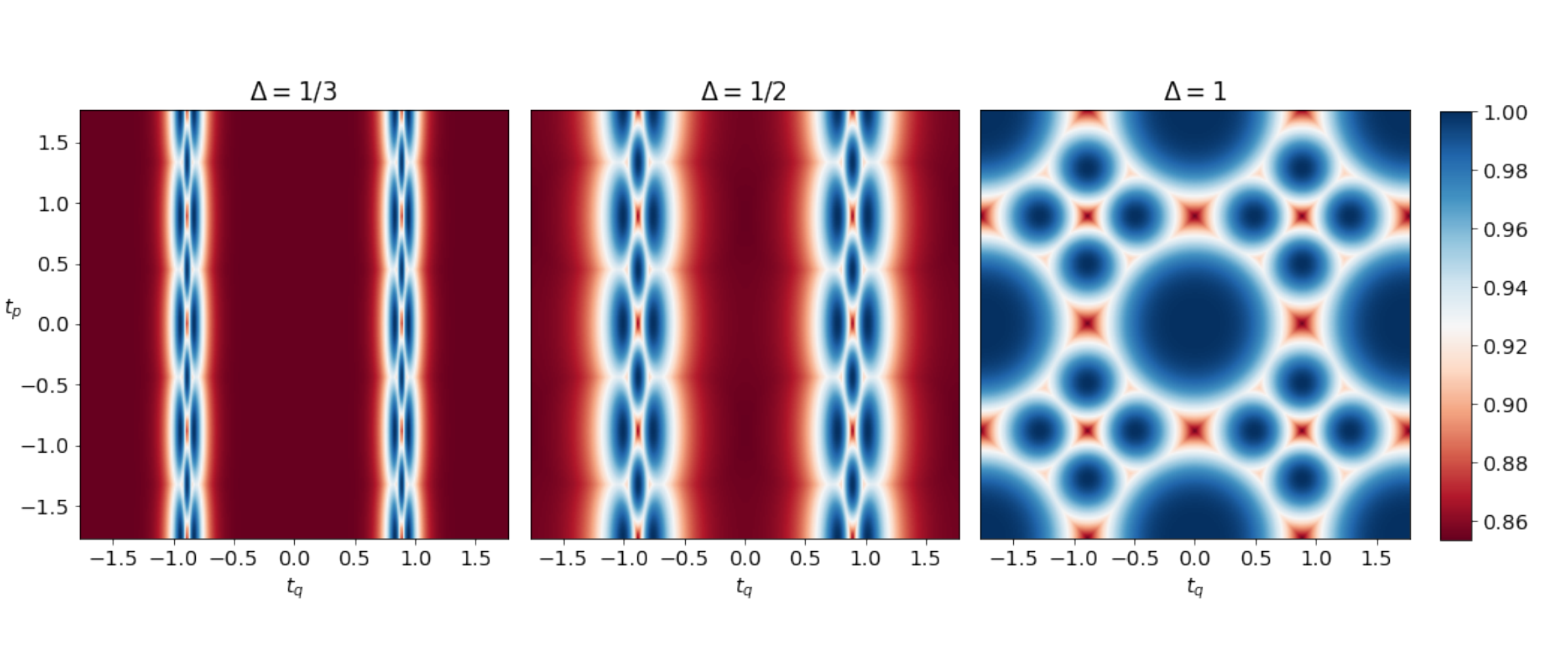}
        \caption{Fidelity of the output state of the error correcting circuit in Fig.~\ref{fig:nonidealgkpinput} with the closest $H$-type magic state, for the various possible measurement outcomes $t_q$ and $t_p$. Note that in the limit $\Delta\to1$ the finitely-squeezed GKP state at the input in Fig.~\ref{fig:nonidealgkpinput} is approximately equivalent to the vacuum state, yielding agreement of panel 3 with Ref.~\cite{baragiola2019}.
        \label{fig:squeezingfidelities}}
    \end{figure*}
    
    The non-ideal GKP state can be defined as~\cite{gottesman2001}
    \begin{align}
        \psi_{0,L(\Delta,\kappa)}(x)=&\bra{x}\ket{0_{\text{GKP}}^{\Delta,\kappa}}\nonumber\\
        \propto&\sum_{s\in\mathbb Z} e^{-2\kappa^2s^2\pi}e^{-(x-2s\sqrt\pi)^2/2\Delta^2}.
    \end{align}
        The output of the circuit of Fig. \ref{fig:nonidealgkpinput} will be a state of the form
    \begin{align}
        \ket{\psi}\propto \hat \Pi_{\text{GKP}}e^{it_q\hat p}e^{-it_p\hat q}\ket{0_{\text{GKP}}^{\Delta,\kappa}},
    \end{align}
    which can be expressed in terms of the coefficients
    \begin{align}
        c_0&=\bra{0_{\text{GKP}}}e^{it_q\hat p}e^{-it_p\hat q}\ket{0_{\text{GKP}}^{\Delta,\kappa}}\nonumber\\
        c_1&=\bra{1_{\text{GKP}}}e^{it_q\hat p}e^{-it_p\hat q}\ket{0_{\text{GKP}}^{\Delta,\kappa}},
    \end{align}
    which can be normalized as
    \begin{align}
        \bar c_0=&\frac{c_0}{\sqrt{c_0^2+c_1^2}}\nonumber\\
        \bar c_1=&\frac{c_1}{\sqrt{c_0^2+c_1^2}}.
    \end{align}

The fidelity of this state with  each of the $\ket{H}$-type states can be calculated in terms of these normalized coefficients. I.e. for each $\ket{H}$-type state $\ket{H}=a_0\ket{0}+a_1\ket{1}$ we calculate the fidelity 
        \begin{align}
            F(\ket{H},\ket \psi)=&\abs{\bra{H}\frac{1}{\sqrt{c_0^2+c_1^2}}(c_0\ket{0}+c_1\ket{1})}^2\nonumber\\
        =&\frac{|a_0c_0+a_1c_1|^2}{\abs{c_0^2+c_1^2}}.
        \end{align}
        
        The probability density function of the measurement outcomes can be calculated, as in \cite{baragiola2019}, as $\pdf(\mathbf t)\propto c_0^2+c_1^2$, which is then normalized over a region periodic in $2\sqrt\pi$ in both $t_q$ and $t_p$.
        The probability of obtaining a state with fidelity higher than a certain threshold $F^*$ can then be calculated numerically by calculating the fidelities for each value of $t_q,t_p$ and integrating over the $\pdf(\mathbf t)$ for those values at which $F>F*$.
        
\begin{figure}
            \centering
            \includegraphics[width=\linewidth]{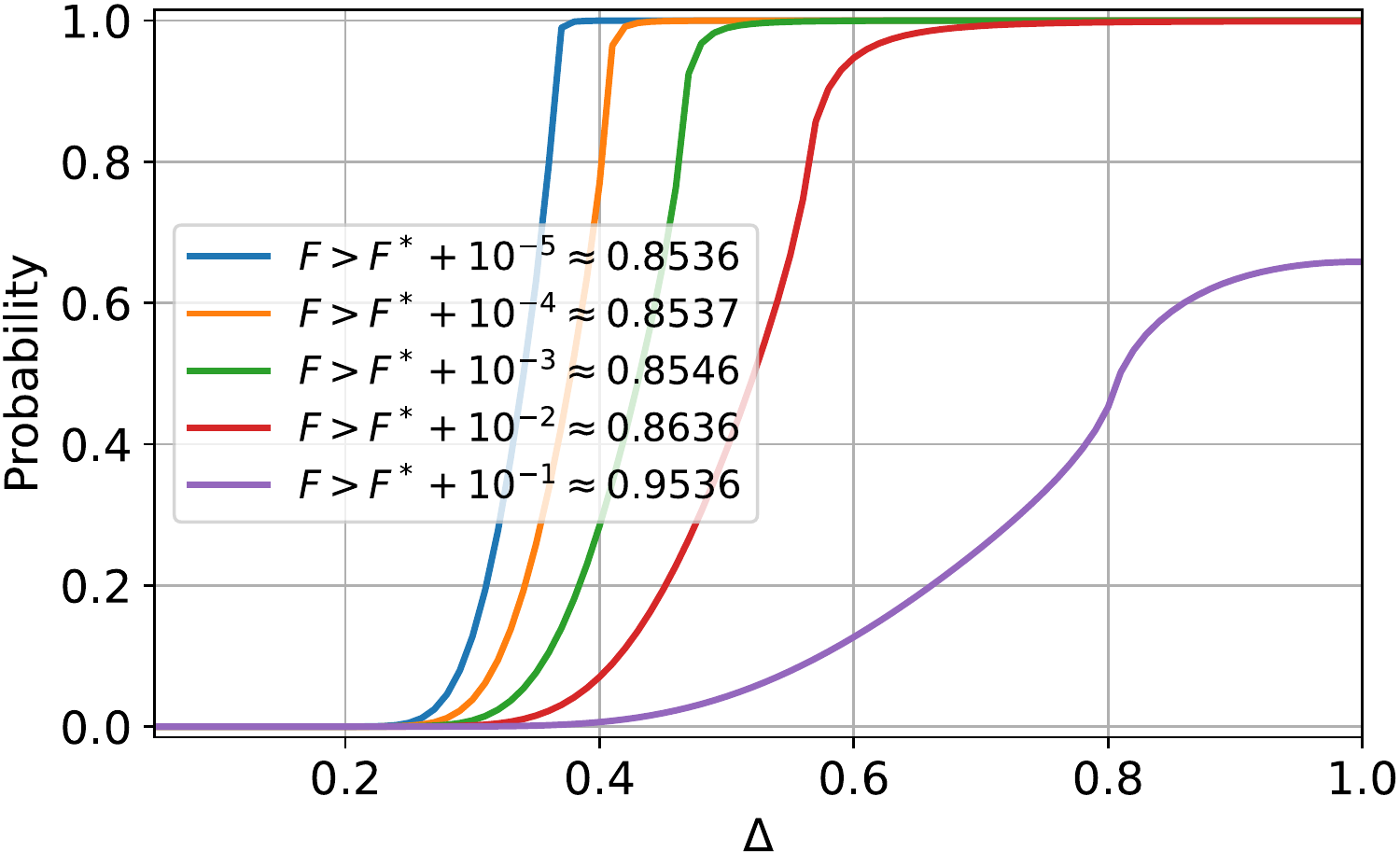}
             \caption{Probability of producing, as output of the circuit in Fig.~\ref{fig:nonidealgkpinput}, a magic state $\ket\psi$ with a given fidelity $F=\abs{\bra{H}\ket{\psi}}^2$ to the nearest target magic state $\ket{H}$ when the additional input state is a GKP state with squeezing level $\Delta=\kappa$. Note that in the other modes we still assume ideal GKP states in logical state 0. Also note that $\Delta\to 1$ is approximately equivalent to the vacuum state. $F^*=\frac{1}{2}(1+\tfrac{1}{\sqrt 2})\approx 0.8536$ is the threshold for magic state distillation \cite{baragiola2019,reichardt2005}.}
            \label{fig:my_label}
        \end{figure}

        In Fig.~\ref{fig:my_label} we plot the probability of obtaining a magic $\ket{H}$-type state in the output of the circuit in Fig.~\ref{fig:nonidealgkpinput} above a given threshold fidelity, for different values of the squeezing parameter $\Delta$.
        
        We see that the squeezing parameter in the auxiliary state of the circuit in Fig.~\ref{fig:nonidealgkpinput} inversely quantifies the resourcefulness of the auxiliary state.

        This result can be understood by interpreting the vacuum as the zero-squeezing limit of a GKP state.        
        The zero-squeezing limit corresponds to setting $\Delta=\kappa=1$. In this case, we obtain from the expression of the finitely-squeezed GKP state in the position representation
        \begin{align}
            \ket{0_{\text{GKP}}^{\Delta,\kappa}} \propto \sum_{s\in\mathbb Z}\int \dd q e^{-2s^2\pi}e^{-(q-2s\sqrt\pi)^2/2}\ket{q}
        \end{align}
    Then, calculating the fidelity with the vacuum state, $\ket{\emptyset}=\pi^{-1/4}\int \dd q e^{-q^2/2}\ket{q}$ gives
    \begin{align}
        \abs{\bra{\emptyset}\ket{0_{\text{GKP}}^{\Delta,\kappa}}}^2=0.999993.
    \end{align}
This high fidelity value explains in particular why, using the zero-squeezing limit of a GKP state, we obtain the plot in Fig.~\ref{fig:squeezingfidelities}, third panel, that is indistinguishable for the naked eye from that of Ref.~\cite{baragiola2019} obtained using input vacuum.

        This allows us to interpret the value $\Delta$ as interpolating from a free state, the ideal 0-logical GKP state, corresponding to infinite squeezing, and with which no magic state can be generated, to a maximally resourceful state, namely the vacuum, corresponding to zero squeezing. 
        
        We can therefore conclude that in the context of this model, realistic GKP states are also resourceful for quantum advantage. The realistic GKP states required for magic state distillation can have any non-infinite squeezing; in other words, there is no threshold required to distill a magic state.

\section{Conclusions}
\label{sec:conclusion}
First, we have demonstrated that circuits with input GKP states acted on with arbitrary displacements and rational~\footnote{Gaussian operations parameterized by irrational symplectic operations cannot in general be simulated with our method. We refer to our previous work~\cite{calcluth2022} which demonstrates that when the symplectic matrix is irrational, the wavefunction of the transformed state corresponds to a periodic distribution which cannot be analytically reduced. Measuring in the position basis of a state which has been transformed by a general irrational symplectic matrix will have a PDF which will give random integer combinations of irrational numbers. Except for specific choices of irrational symplectic matrices, the measurement values will be randomly selected from a set dense on the real number line. } symplectic operations, and measured with homodyne detection are classically efficiently simulatable.
This result extends the classes of circuits previously known to be simulatable and can be understood as a CV analogue to the Gottesman-Knill theorem~\cite{gottesman1997, gottesman1999,nielsen2000}. The Gottesman-Knill theorem provides a method to simulate circuits involving qubits initialized in ideal input qubit stabilizer states acted on by Clifford operations and measured in the computational basis. Meanwhile, our result provides a method to simulate ideal GKP states acted on by Gaussian operations and measured with homodyne detection.

Second, this result in combination to those of Ref.~\cite{baragiola2019} leads to the counter-intuitive interpretation of the vacuum, or realistic GKP states with any finite squeezing, as a resource for universality. 
Here we can draw an analogy to DV magic state distillation~\cite{bravyi2005}, where it is known that ``noisy'' pure states that are close to (but not exactly) the points corresponding to the stabilizer states on the Bloch sphere act as a resource for universal QC~\cite{reichardt2005}.  Similarly, in the circuits we have considered, introducing noise in the form of vacuum  or realistic GKP states  promotes the circuit class we consider to universality by allowing one to produce and distill magic states.

The question of whether realistic GKP states in all modes (possibly with different squeezing levels), combined with Gaussian operations, yield a simulatable or universal model is still open. Our analysis in Sec.~\ref{sec:realistic-gkp-states-resource}
, showing that a combination of ideal and realistic GKP states yields a universal model, can be seen as a first attempt to provide an answer to this question.
Therein, squeezing quantifies inversely the resourcefulness of finitely-squeezed GKP states, when these are combined with infinitely-squeezed GKP states, in terms of their ability of producing output GKP magic states. This leads us to speculate that, even in a more realistic model with input highly-squeezed GKP states, the vacuum will retain its character as a resource, boosting the magic content at the output of the circuit.

Our work also opens the question as to whether the methodology introduced to compute the PDF, based on imposing stabilizer conditions, can be also used for other types of circuits, for which input states admit a stabilizer representation. 

\section{Acknowledgments}
We acknowledge useful discussions with Laura Garc\'ia-\'Alvarez and Ben Q. Baragiola.
G. F. and C. C. acknowledge support from the VR (Swedish Research Council) Grant QuACVA and the Wallenberg Center for Quantum Technology (WACQT). 

\appendix
\section{The set of rational matrices is dense in the reals}
\label{appendix:density-proof}
        In this appendix, we will prove that
        $\text{Sp}(2n,\mathbb Q)$ is dense on $\text{Sp}(2n,\mathbb R)$. This is equivalent~\cite{croom2016} to showing that the closure of the rational symplectic group $\text{cl}(\text{Sp}(2n,\mathbb Q))$ is the real symplectic group $\text{Sp}(2n,\mathbb R)$, i.e. $\text{cl}(\text{Sp}(2n,\mathbb Q))=\text{Sp}(2n,\mathbb R)$.
        
    We provide an overview of the steps of this proof taken from Ref.~\cite{stackexchangedensity}.
      First note that the symplectic group defined over any field $\text{Sp}(2n,\mathbb F)$ is generated by the set of symplectic transvections, $H_{\mathbb F}$~\cite{artin1988}. This set consists of maps $f_{\alpha,\vec u}$ with $\alpha \in \mathbb{F}$ and $\vec u \in \mathbb{F}^2$ which transforms any arbitrary vector $\vec v \in \mathbb{F}^{2n}$ as
        \begin{align}
         f_{\alpha,\vec u}(\vec v) = \vec v + \alpha B(\vec v,\vec u) \vec u
        \end{align}
        where $B$ is the alternating bilinear form~\cite{stackexchangedensity,weyl1946}.
        
        The set of generators of the rational symplectic group can therefore be written as
        \begin{align}
            H_{\mathbb Q}=\{f_{\alpha,\vec u} : \alpha\in \mathbb Q, \vec u\in \mathbb Q^{2n}\},
        \end{align}
        while the set of generators of the real symplectic group can be written as
        \begin{align}
            H_{\mathbb R}=\{f_{\alpha,\vec u} : \alpha\in \mathbb R, \vec u\in \mathbb R^{2n}\}.
        \end{align}
        For any chosen generator in the set of generators for the real symplectic group, $f_{\alpha,\vec u} \in  H_{\mathbb R}$, it is possible to find an arbitrarily close generator from the set of generators of the rational symplectic group $f_{ \alpha',\vec u'} \in  H_{\mathbb Q}$.
        We can demonstrate this by evaluating the norm of the difference of these generators~\cite{stackexchangedensity} and showing that it is possible to find for any $f_{ \alpha,\vec u}$ and $f_{ \alpha',\vec u'}$ a norm such that
         \begin{align}
            ||f_{ \alpha',\vec u'}(\vec v)-f_{ \alpha,\vec u}(\vec v)||=&|| \alpha' B(\vec v,\vec u') \vec u'- \alpha B(\vec v,\vec u) \vec u||\le  \epsilon.
         \end{align}

        Choosing $\vec u'=\vec u+\vec u_\epsilon$ and  $\alpha'=\alpha+\alpha_{\epsilon}$ we can make use of the triangle inequality \cite{pedoe1988} to find
         
        \begin{align}
            &||f_{ \alpha',\vec u'}(\vec v)-f_{ \alpha,\vec u}(\vec v)||\nonumber\\
            =&|| \alpha' B(\vec v,\vec u+\vec u_\epsilon) (\vec u+\vec u_\epsilon)- \alpha B(\vec v,\vec u) \vec u||\nonumber\\
            =&|| (\alpha+\alpha_\epsilon) (B(\vec v,\vec u_\epsilon)+B(\vec v,\vec u)) (\vec u+\vec u_\epsilon)- \alpha B(\vec v,\vec u) \vec u|| \nonumber\\
            \le&|\alpha B(\vec v,\vec u_\epsilon)| \cdot ||  \vec u||+|\alpha B(\vec v,\vec u_\epsilon)| \cdot ||\vec u_\epsilon|| \nonumber\\
            &+|\alpha B(\vec v,\vec u)|\cdot||\vec u_\epsilon||+|\alpha_\epsilon B(\vec v,\vec u_\epsilon)|\cdot ||\vec u||\nonumber \\
            &+|\alpha_\epsilon B(\vec v,\vec u)|\cdot ||\vec u||+|\alpha_\epsilon B(\vec v,\vec u_\epsilon)| \cdot ||\vec u_\epsilon|| \nonumber\\
            &+|\alpha_\epsilon B(\vec v,\vec u)| \cdot ||\vec u_\epsilon ||.
         \end{align}
        Note that we can write $\vec u_{\epsilon}=\epsilon_{u} \vec u_{\epsilon}^{\mathbbm 1}$ where $\vec u_{\epsilon}^{\mathbbm 1}$ is the unit vector containing the direction of $\vec u_{\epsilon}$, i.e. $|\vec u_{\epsilon}^{\mathbbm 1}|=1$, and the magnitude $\epsilon_{\vec u} \in \mathbb R$ is small $\epsilon_{\vec u}\ll 1$.
        
        This allows us to write 
        \begin{align}
            B(\vec v,\vec u_{\epsilon})=\epsilon_{u}B(\vec v,\vec u_{\epsilon}^{\mathbbm 1}).
        \end{align}
        Therefore, for any chosen $7\epsilon\in \mathbb R$, we can ensure that the distance is less than $7\epsilon$ by ensuring each term is smaller than $\epsilon$, i.e.
        \begin{align}
            &|\alpha B(\vec v,\epsilon_{u}\vec u_{\epsilon}^{\mathbbm 1})| \cdot ||  \vec u||\le \epsilon\\
            &|\alpha B(\vec v,\epsilon_{u}\vec u_{\epsilon}^{\mathbbm 1})| \cdot ||\epsilon_{u}\vec u_{\epsilon}^{\mathbbm 1}|| \le \epsilon\\
            &|\alpha B(\vec v,\vec u)|\cdot||\epsilon_{u}\vec u_{\epsilon}^{\mathbbm 1}||\le \epsilon\\
            &|\alpha_\epsilon B(\vec v,\epsilon_{u}\vec u_{\epsilon}^{\mathbbm 1})|\cdot ||\vec u||\le \epsilon\\
            &|\alpha_\epsilon B(\vec v,\vec u)|\cdot ||\vec u||\le \epsilon\\
            &|\alpha_\epsilon B(\vec v,\epsilon_{u}\vec u_{\epsilon}^{\mathbbm 1})| \cdot ||\epsilon_{u}\vec u_{\epsilon}^{\mathbbm 1}||\le \epsilon\\
            &|\alpha_\epsilon B(\vec v,\vec u)| \cdot ||\epsilon_{u}\vec u_{\epsilon}^{\mathbbm 1} ||\le \epsilon
        \end{align}
        which is equivalent to
        \begin{align}
            &|\epsilon_{u}\alpha B(\vec v,\vec u_{\epsilon}^{\mathbbm 1})| \cdot ||  \vec u||\le \epsilon\\
            &|\alpha\epsilon_{u} B(\vec v,\vec u_{\epsilon}^{\mathbbm 1})| \cdot \epsilon_{u} \le \epsilon\\
            &|\alpha B(\vec v,\vec u)|\epsilon_{u}\le \epsilon\\
            &|\alpha_\epsilon \epsilon_{u} B(\vec v,\vec u_{\epsilon}^{\mathbbm 1})|\cdot ||\vec u||\le \epsilon\\
            &|\alpha_\epsilon B(\vec v,\vec u)|\cdot ||\vec u||\le \epsilon\\
            &|\alpha_\epsilon \epsilon_{u} B(\vec v,\vec u_{\epsilon}^{\mathbbm 1})| \cdot \epsilon_{u}\le \epsilon\\
            &|\alpha_\epsilon B(\vec v,\vec u)| \cdot \epsilon_{u}\le \epsilon.
        \end{align}
        These can be rearranged into conditions
        \begin{align}
            &\epsilon_u\le \epsilon/(|\alpha B(\vec v,\vec u_{\epsilon}^{\mathbbm 1})|\cdot ||  \vec u||)\\
            &\epsilon_u^2\le \epsilon/|\alpha B(\vec v,\vec u_{\epsilon}^{\mathbbm 1})|\\
            &\epsilon_{u}\le \epsilon/|\alpha B(\vec v,\vec u)|\\
            &\alpha_\epsilon \epsilon_{u}\le \epsilon/(| B(\vec v,\vec u_{\epsilon}^{\mathbbm 1})|\cdot ||\vec u||)\\
            &\alpha_\epsilon \le \epsilon/(| B(\vec v,\vec u)|\cdot ||\vec u||)\\
            &\alpha_\epsilon \epsilon_{u}^2\le \epsilon/ | B(\vec v,\vec u_{\epsilon}^{\mathbbm 1})|\\
            &\alpha_\epsilon  \epsilon_{u}\le \epsilon/|B(\vec v,\vec u)|
        \end{align}
        and can be further simplified to the conditions
        \begin{align}
            &\epsilon_u\le \epsilon/(|\alpha B(\vec v,\vec u_{\epsilon}^{\mathbbm 1})|\cdot ||  \vec u||)\\
            &\epsilon_u\le \sqrt{\epsilon/|\alpha B(\vec v,\vec u_{\epsilon}^{\mathbbm 1})|}\\
            &\epsilon_{u}\le \epsilon/|\alpha B(\vec v,\vec u)|\\
            &\alpha_\epsilon \le 1/\alpha\\
            &\alpha_\epsilon \le \epsilon/(| B(\vec v,\vec u)|\cdot ||\vec u||).
        \end{align}
        For any $\alpha,\vec u, \vec v$ it is always possible to find $\alpha',\vec u'$ for which $\alpha_\epsilon,\epsilon_{u}\in\mathbb R$ are arbitrarily close to $0$ such that all these inequalities hold. This follows from the fact that the rational numbers are dense on the reals~\cite{trench2003}.
        
    We can therefore say that $H_{\mathbb Q}$ is dense on the set  $H_{\mathbb R}$, which can be expressed in terms of the closure of the set of rational generators $\text{cl}(H_{\mathbb Q})=H_{\mathbb R}$.
        
        Furthermore, we know that $H_{\mathbb Q} \subseteq \text{Sp}(2n,\mathbb Q)$, which implies that \cite{croom2016} $\text{cl}(H_{\mathbb Q}) \subseteq \text{cl}(\text{Sp}(2n,\mathbb Q))$ and hence the generators of $H_{\mathbb R}$ are all members of the closure of the rational symplectic group, i.e. $H_{\mathbb R} \subseteq \text{cl}(\text{Sp}(2n,\mathbb Q))$.
         Since all the generators of $\text{Sp}(2n,\mathbb R)$ are members of $\text{cl}(\text{Sp}(2n,\mathbb Q))$, then $\text{Sp}(2n,\mathbb R) \subseteq \text{cl}(\text{Sp}(2n,\mathbb Q))$.
        
        Finally, using the fact that $\text{Sp}(2n,\mathbb Q)\subseteq \text{Sp}(2n,\mathbb R)$, we have $\text{cl}(\text{Sp}(2n,\mathbb Q))\subseteq \text{Sp}(2n,\mathbb R)$. This means that $\text{cl}(\text{Sp}(2n,\mathbb Q))=\text{Sp}(2n,\mathbb R)$ and the symplectic group over the rationals is dense on the symplectic group over the reals.
    
\section{Solution to the constrained linear equation}
\label{sec:appendix-solution-constrained-eq}
In this appendix, we solve the constrained equation introduced in Sec.~\ref{sec:efficient-algorithm}, which provides the solution for the set of allowed points of the PDF. Specifically, we find the solution of Eq.~(\ref{eq:condition2}) given the constraint of Eq.~(\ref{eq:condition1}). I.e., 
\begin{align}
    \label{eq:appendix-constrained}
    & \sqrt\pi \vec l^T \vec x-\frac 1 2 \pi \vec l^T AB^T\vec l-\sqrt\pi \vec l\cdot \vec c=0\mod 2\pi \nonumber \\
    & \quad \text{s.t.} \quad \begin{matrix} (A^T\vec l)_{k}=&0\mod 1\\
    (B^T\vec l)_{k}=&0\mod 2\end{matrix}
\end{align}
To solve this equation, we begin by identifying a method to evaluate the possible values of the vector $\vec l$.

First, in Sec.~\ref{sec:appendix-identifying-l}, we express the constraint as an overdetermined system of linear equations, that has solutions dependent on a projection matrix $1-SS^+$. In Sec.~\ref{sec:appendix:projector-rational}, we demonstrate that this projection matrix is rational, given that the symplectic matrix is rational. Together, these results allow us to provide the solutions to $\vec l$ in Sec.~\ref{sec:appendix-evaluation-l}.
Then, in Sec.~\ref{sec:appendix-linear} we demonstrate that given the solutions of $\vec l$, we can express the constrained equation in Eq.~(\ref{eq:appendix-constrained}) as set of unconstrained linear equations. Finally, in Sec.~\ref{eq:appendix-solutions}, we provide the solution of these unconstrained equations, which are also the solutions to the constrained equation given in Eq.~(\ref{eq:appendix-constrained}).

\subsection{Expressing the constraint as a linear system of equations}
\label{sec:appendix-identifying-l}
We first identify a method to express the constraining terms defined in Eq.~(\ref{eq:appendix-constrained}) as a system of overdetermined linear equations. We demonstrate that the solutions of $\vec l$ can be expressed in terms of the pseudo-inverse of a matrix $S$, which is dependent on $A$ and $B$, and a new vector $\vec b$, which will be solved in the following subsections. To begin, we combine the constraining terms into one equation of the form
\begin{align}
    \mqty(A^T\\ B^T)\vec l=\mqty(0\mod 1\\ \vdots \\ 0\mod 1\\0\mod 2\\ \vdots \\ 0\mod 2).
\end{align}
We now introduce a matrix $S$ which is defined in terms of the two matrices $A$ and $B$ as
\begin{align}
    S=\mqty(A^T\\ \frac 1 2B^T)=\mqty(1&0\\0&\frac 1 2 )\bar S,
\end{align}
where we also introduce the matrix $\bar S$, which is the transpose of the first $n$ rows of the symplectic matrix $M$, i.e.,
\begin{align}
    \label{eq:sbardef}
    \bar S=\mqty(A^T\\B^T).
\end{align}
We introduce the vector $\vec b$ which is a $2n$-vector of integers. This allows us to express the constraint on $\vec l$ as
\begin{align}
    &S\vec l=\vec b.
\end{align}
This gives an overdetermined system of linear equations and does not necessarily always have a solution. Whether the system has solutions or not depends on which integers are chosen in $\vec b$.

The columns of $S$ are linearly independent. This can be seen by considering the fact that the determinant of the symplectic matrix is $\det M=1$ which means that it has linearly independent rows~\cite{greub1975}. Hence, the matrix $\bar S$ will have linearly independent columns. Furthermore, the matrix which converts $\bar S$ to $S$ is a full rank $2n\times 2n$ matrix. Hence, the rank of $S$ will be the same as the rank of $\bar S$, i.e. it will have rank $n$ which means the columns must be linearly independent~\cite{roman2007,stackexchangesmith}.

We can express the solutions of $\vec l$ in terms of the Moore-Penrose pseudoinverse~\cite{moore1920,penrose1955,ben2003}, which is a generalization of the matrix inverse. For any matrix $S$ there exists a pseudoinverse $S^+$ even if the matrix does not have a true inverse. A $2n\times n$ rectangular matrix $S$ with linearly independent columns has rank $n$~\cite{roman2007}. The pseudoinverse $S^+$ is defined such that $S^+S=\mathbbm 1$. The pseudoinverse of $S$ can be found in terms of the pseudoinverse of the rank $n$ matrix $\bar S$ as~\cite{golub1996}
\begin{align}
    S^+=\bar S^+\mqty(1&0\\0&1/2)^+=\bar S^+\mqty(1&0\\0&2)
\end{align}
where we have used the fact that the pseudoinverse of a non-singular matrix is equal to its inverse~\cite{ben2003}.

This gives potential solutions of $\vec l$ in the form
\begin{align}
    \label{eq:possible-sols-l}
    \vec l=S^+\vec b
\end{align}
but if this system is unsolvable for a given $\vec b$ then the pseudoinverse will not provide a valid solution to $\vec l$. For it to be valid it must satisfy the original equation~\cite{ben2003}
\begin{align}
    S\vec l=SS^+\vec b={\vec b}
\end{align}
which gives the constraint on the integers $\vec b$ as
\begin{align}
    \label{eq:appendix-vecb-constraint}
    SS^+\vec b=\vec b.
\end{align}
We can write this constraint as
\begin{align}
    (SS^+-1)\vec b=0 \quad     \implies \quad (1-SS^+)\vec b=0. \label{eq:appendix-projector}
\end{align}
Finding the values of $\vec b$ which satisfy this equation also informs us of all the possible choices of $\vec l$ which satisfy the constraining equation. This is equivalent to finding the eigenvectors of the projection matrix $1-SS^+$ which have eigenvalues equal to zero. To solve this equation, we will demonstrate that the projection matrix $1-SS^+$ has a convenient eigenvalue decomposition, provided that it is a rational matrix. We first prove that this matrix is rational in the following section, given that the symplectic matrix is rational, and then we will proceed to find its eigenvectors.

\subsection{The projector is rational}
\label{sec:appendix:projector-rational}
In this subsection, we will analyze the $2n\times 2n$ projection matrix $1-SS^+$ and demonstrate that it contains all rational elements. We then use the expression to find the pseudoinverse of a matrix with linearly independent columns, to identify the pseudoinverse of $\bar S$ as~\cite{ben2003}
\begin{align}
    \label{eq:sbarpseudo}
    \bar S^+=(\bar S^T\bar S)^{-1}\bar S^T.
\end{align}

Note that $1-SS^+$ will be rational if $SS^+$ is rational. Inspecting
\begin{align}
    \bar S^+=(AA^T+BB^T)^{-1}\mqty(A & B)
\end{align}
we can see that as long as $A,B$ are rational, the matrix $(AA^T+BB^T)$ will be rational. The inverse of a rational matrix will also be rational, and so $\bar S^+$ will also be rational. Therefore we know $S^+$ is rational. This also implies that $1-SS^+$ is a matrix of rational elements.

\subsection{Evaluation of the allowed parameters provided by the constraint}
\label{sec:appendix-evaluation-l}
As shown in the previous subsection, the matrix ${1-SS^+}$ consists of all rational elements. In this subsection, we will demonstrate that it has an eigenvector decomposition of the form
\begin{align}
    1-SS^+=V\mqty(0&0\\0&\mathbbm 1)V^{-1}
\end{align}
where $V$ is a unimodular matrix, also known as a unit matrix~\cite{ben2003}. The definition of a unimodular matrix is one that contains all integers and has determinant $1$~\cite{newman1972}. This decomposition then be used to find the solutions of $\vec b$ in Eq.~(\ref{eq:appendix-projector}) and therefore also $\vec l$ in Eq.~(\ref{eq:possible-sols-l}).

To find such $V$ for a given matrix $1-SS^+$ we can first find the Smith decomposition of the matrix $\sigma S$. We use the integer $\sigma$ to multiply every element of matrix $S$ to an integer. The integer $\sigma$ can be found to be the lowest common multiple of all of the denominators of $S$. The Smith decomposition is given by
\begin{align}
    \sigma S=VDU \quad \implies \quad S=\sigma^{-1}VDU
\end{align}
where $V$ is a $2n\times 2n$ unimodular matrix, $U$ is a $n\times n$ unimodular matrix. $D$ is a diagonal $2n\times n$ matrix which has the same rank as $S$, which has rank $n$. The Smith decomposition algorithm will order the diagonal elements of $D$ in descending order. We can therefore assume that $D$ has $n$ non-zero entries along the diagonal. The remaining entries in the matrix $D$ will be $0$.

Furthermore, we can identify the pseudoinverse of $S$ as
\begin{align}
    S^+=\sigma U^{-1}D^+V^{-1},
\end{align}
where we have used that the pseudoinverse of the product of two matrices $AB$ is $(AB)^+=B^+A^+$~\cite{greville1966}.
This gives a convenient expression for $SS^+$
\begin{align}
    \label{eq:unimodulardecomposition}
    SS^+=VDD^+V^{-1}
\end{align}
and the projector
\begin{align}
    1-SS^+=1-VDD^+V^{-1}=V(1-DD^+)V^{-1}.
\end{align}
As $D$ is a matrix of integer entries along the diagonal, its pseudoinverse, $D^+$, can be found by taking the inverse of each non-zero element along the diagonal and then transposing~\cite{ben2003}. Therefore we can find $DD^+$ by inspecting its form,
\begin{align}
    \mqty(D_{1,1}&\dots&0\\0&\ddots&0\\
    0&\dots&D_{n,n}\\
    0&\dots &0\\
    0&\vdots &0\\
    0&\dots & 0)\mqty(D_{1,1}^{-1}&\dots&0&0&\dots&0\\0&\ddots&0&0&\dots&0\\
    0&\dots&D_{n,n}^{-1}&0&\dots&0),
\end{align}
which means that
\begin{align}
    DD^+ =\mqty(\mathbbm 1&0\\0&0).
\end{align}
From this, we can immediately identify
\begin{align}
    \label{eq:appendix-DDproj}
    1-DD^+=\mqty(0&0\\0&\mathbbm 1)
\end{align}
and we can express the projector as
\begin{align}
    1-SS^+=1-VDD^+V^{-1}=V\mqty(0&0\\0&\mathbbm 1)V^{-1}
\end{align}
as we had anticipated.
This form is an eigenvalue decomposition of the matrix $1-SS^+$.
The integer eigenvectors of $1-SS^+$ are given as the columns of $V$. The first $n$ columns correspond to eigenvectors with eigenvalue $0$ and the remaining $n$ columns correspond to eigenvectors with eigenvalue $1$. We can therefore construct any integer eigenvector with eigenvalue $0$ as \cite{stackexchangesmith}
\begin{align}
    \label{eq:appendix-vecb}
    \vec b=V\mqty(\mathbbm 1\\0)\vec m.
\end{align}

Note that in general the decomposition matrices $U,V$ of the Smith decomposition are not necessarily unique. However, the complete set of eigenvectors $\vec b$ will be the same regardless of which decomposition matrix $V$ is found~\cite{stackexchangesmith}.

The allowed values of $\vec l$ can therefore be calculated in terms of the values of $\vec b$ using Eq.~(\ref{eq:possible-sols-l}), which can be expressed in terms of the $n$-vector $\vec m$ as
\begin{align}
    \label{eq:appendix-l-solution}
    \vec l=S^+\vec b=S^+V\mqty(\mathbbm 1\\0)\vec m=R\vec m
\end{align}
where we have introduced the $n\times n$ matrix $R$ as
\begin{align}
    \label{eq:appendix-R-def}
    R=S^+V\mqty(\mathbbm 1\\0).
\end{align}
Given these solutions for the vector $\vec l$, we can solve the constrained equation, given in Eq.~(\ref{eq:appendix-constrained}), to find the allowed values of $\vec x$.

\subsection{Expressing the constrained equation as a set of linear equations}
\label{sec:appendix-linear}

We then would like to solve the
equation
\begin{align}
    \sqrt\pi \vec l^T \vec x-\frac 1 2 \pi \vec l^T AB^T\vec l-\sqrt\pi \vec l\cdot \vec c=0\mod 2\pi
\end{align}
for which we can, without loss of generality, set $\vec c=0$ (because we can consider a different $\vec c$ to be a change of variables in $\vec x$) and it becomes
\begin{align}
    \label{eq:appendix-constrained-linear-eq}
    \vec l^T\left(\frac{1}{\sqrt\pi} \vec x-\frac 1 2  AB^T\vec l\right)=0\mod 2.
\end{align}

We consider the term
\begin{align}
    \label{eq:appendix-quadratic-T}
    \frac 1 2 \vec l^TAB^T\vec l=&\vec m^T T\vec m
\end{align}
where we have defined the $n\times n$ matrix
\begin{align}
    \label{eq:appendix-T}
    T=&\frac 1 2 R^TAB^TR\nonumber\\
    =&\frac 1 2\mqty(\mathbbm 1&0)V^{T}(S^+)^TAB^TS^+V\mqty(\mathbbm 1\\0).
\end{align}
The matrix $T$ will always give integer values. For proof of this consider the following. We know from Eq.~(\ref{eq:appendix-vecb-constraint}) and Eq.~(\ref{eq:appendix-vecb}) that
\begin{align}
    SS^+V\mqty(\mathbbm 1\\0)\vec m=V\mqty(\mathbbm 1\\0)\vec m
\end{align}
which must be true for all integer vectors $\vec m$. As a consequence we have
\begin{align}
    &SS^+V\mqty(\mathbbm 1\\0)=V\mqty(\mathbbm 1\\0)\nonumber\\
    \implies & \mqty(A^T\\ \frac 1 2 B^T)S^+V\mqty(\mathbbm 1\\0)=V\mqty(\mathbbm 1\\0).
    \end{align}
We know that
\begin{align}
    S^+V\mqty(\mathbbm 1\\0)
\end{align}
is an $n\times n$ matrix so we must have
\begin{align}
    A^TS^+V\mqty(\mathbbm 1\\0)=V^{(11)} \nonumber\\
    \frac 1 2 B^TS^+V\mqty(\mathbbm 1\\0)=V^{(21)}.
\end{align}
This means that from Eq.~(\ref{eq:appendix-T}) the matrix $T$ can be succinctly written as
\begin{align}
    T=&V^{(11)T}V^{(21)}.
\end{align}
The matrix $V$ is unimodular meaning that it consists of all integer elements. The block matrices $V^{(21)},V^{(11)}$ must also be integer and the multiplication of two integer matrices is also integer. Hence, $T$ is an integer matrix.

We can now solve   Eq.~(\ref{eq:appendix-constrained-linear-eq}) which constrains the values of $\vec x$, which can be written in terms of Eq.~(\ref{eq:appendix-l-solution}) and Eq.~(\ref{eq:appendix-quadratic-T}) as
\begin{align}
    \frac{1}{\sqrt\pi}\vec m^T R^T\vec x-\vec m^T T \vec m=0\mod 2.
\end{align}
This must be true for any chosen $\vec m$. We introduce the length-$n$ basis vector $\vec e^{(j)}$ which has zero in all elements, except at element $j$ for which it is $1$,
\begin{align}
    \label{eq:appendix-basis-vector}
    \vec e^{(j)}=(0_1,\dots,0_{j-1},1_j,0_{j+1},\dots,0_n)^T.
\end{align}
Choosing $\vec m=m_j \vec e^{(j)}$, for any integer $m_j\in\mathbb Z$, gives an equation of the form
\begin{align}
    \frac{1}{\sqrt\pi}m_j(R^T\vec x)_j-m_j^2T_{jj}=0\mod 2.
\end{align}
The vector $\vec m=m_j\vec e^{(j)}$ will produce constraints for different choices of $m_j\in\{1,2,3,\dots\}$ as
\begin{align}
    \frac{1}{\sqrt\pi}(R^T\vec x)_j-T_{jj}&=0\mod 2\\
    2\frac{1}{\sqrt\pi}(R^T\vec x)_j-4T_{jj}&=0\mod 2\\
    3\frac{1}{\sqrt\pi}(R^T\vec x)_j-9T_{jj}&=0\mod 2,
\end{align}
continuing for all integers $m_j$.
We know that $T_{jj}$ is an integer and so we inspect two cases. In the first case we consider when $T_{jj}$ is an even integer. These constraints can then always be simplified to
\begin{align}
m_j\frac{1}{\sqrt\pi}(R^T\vec x)_j=0\mod 2.
\end{align}
Then using the fact that this must hold for any choice of $m_j$ we identify that any integer $m_j$ multiplied by $\frac{1}{\sqrt\pi}(R^T\vec x)_j$ is an even integer. This means that for even $T_{jj}$ we have
\begin{align}
    \frac{1}{\sqrt\pi}(R^T\vec x)_j=0\mod 2.
\end{align}
In the second case, for which $T_{jj}$ is odd and so ${T_{j,j} \mod 2=1}$, the constraints can be simplified to
\begin{align}
    \frac{1}{\sqrt\pi}(R^T\vec x)_j-1=0\mod 2\\
    2\frac{1}{\sqrt\pi}(R^T\vec x)_j=0\mod 2\\
    3\frac{1}{\sqrt\pi}(R^T\vec x)_j-1=0\mod 2
\end{align}
which will be satisfied for all choices of $m_j$ if and only if $\frac{1}{\sqrt\pi}(R^T\vec x)_j$ is an odd number. Hence, for odd $T_{jj}$ we can write
\begin{align}
    \frac{1}{\sqrt\pi}(R^T\vec x)_j=1\mod 2.
\end{align}
Combining these two cases we can express the two relations, which depend on whether $T_{jj}$ is odd i.e. $T_{jj}\mod 2=1$ or even i.e. $T_{jj}\mod 2=0$, as
\begin{align}
    \frac{1}{\sqrt\pi}(R^T\vec x)_j=T_{j,j} \mod 2.
\end{align}

We can also attempt to select for combinations of these basis vectors. For example, we can choose $\vec m=m_i\vec e^{(i)}+m_j\vec e^{(j)}$ for different integers $m_i,m_j\in\mathbb Z$. These will give constraints of the form
\begin{widetext}
\begin{align}
    \frac{1}{\sqrt\pi}m_i(R^T\vec x)_i+\frac{1}{\sqrt\pi}m_j(R^T\vec x)_j-m_j^2T_{jj}-m_i^2T_{ii}-2m_im_jT_{ij}=0\mod 2
\end{align}
\end{widetext}
but because we know that every element $T_{i,j}$ is an integer, this is equivalent to linear combinations of the constraints with a single $m_j\ne 0$. We already know that the constraints with single $m_j\ne 0$ are satisfied and so adding combinations of such constraints do not constraint the allowed values of $\vec x$ any further.

A valid solution can be found by solving
\begin{align}
    \label{eq:appendix-x-constraints}
    \frac{1}{\sqrt\pi}R^T\vec x=\vec t\mod 2
\end{align}
where $\vec t$ is an integer vector of the diagonal elements of $T$. 
Note that $R^T$ is a $n\times n$ matrix given by
\begin{align}
    R^T=\mqty(\mathbbm 1&0)V^{T}(S^+)^T.
\end{align}
This system of equations will have infinite solutions. However, if $R^T$ is invertible then the system of equations can be solved by applying the inverse of $R^T$ to the left of both sides of the equation.
\subsection{Solutions of the constrained equation}
\label{eq:appendix-solutions}
To solve the set of linear equations we need to find the inverse of $R$. We first claim that the pseudoinverse is the inverse of $R$.  
$R$ is given in Eq.~(\ref{eq:appendix-R-def}) so its pseudoinverse is
\begin{align}
    R^+=\mqty(\mathbbm 1&0)V^{-1}S.
\end{align}

Now, we can check that $R^+R=1$ and $RR^+=1$. If this is true then we will know that $R$ is invertible and $R^+=R^{-1}$. First we see that 
\begin{align}
    R^+R=\mqty(\mathbbm 1&0)V^{-1}SS^+V\mqty(\mathbbm 1\\0)
\end{align}
and use Eq.~(\ref{eq:unimodulardecomposition}) and Eq.~(\ref{eq:appendix-DDproj}) to write
\begin{align}
    R^+R=&\mqty(\mathbbm 1&0)DD^+\mqty(0\\\mathbbm 1)\nonumber\\
    =&\mqty(\mathbbm 1&0)\mqty(\mathbbm 1 &0\\0&0)\mqty(\mathbbm 1\\0)\nonumber\\
    =&\mqty(\mathbbm 1&0)\mqty(\mathbbm 1\\0)\nonumber\\
    =&\mathbbm 1.
\end{align}
Furthermore we can check $RR^+$ is equal to the identity
\begin{align}
    RR^+=&S^+V\mqty(\mathbbm 1\\0)\mqty(\mathbbm 1&0)V^{-1}S\nonumber\\
    =&S^+V\mqty(\mathbbm 1&0\\0&0)V^{-1}S.
\end{align}
This time we replace the matrix with $DD^+$, using Eq.~(\ref{eq:unimodulardecomposition}) to find 
\begin{align}
    RR^+=&S^+VDD^+V^{-1}S\nonumber\\
    =&S^+SS^+S\nonumber\\
    =&\mathbbm 1
\end{align}
where we have used that $S^+S=\mathbbm 1$.

This means that $RR^+=R^+R=\mathbbm 1$ which implies that $R^{-1}=R^+$. Hence, we can write the inverse of $R$ as
\begin{align}
    R^{-1}=\mqty(\mathbbm 1 &0)V^{-1}S
\end{align}
and
\begin{align}
    R^{-T}=S^TV^{-T}\mqty(\mathbbm 1 \\0).
\end{align}

Finally, we can invert Eq.~(\ref{eq:appendix-x-constraints}) to identify the solutions to the constrained linear equation as
\begin{align}
    \label{eq:appendix-solutions-nonzero}
     \vec x=\sqrt\pi R^{-T}(\vec t+2\vec m).
\end{align}\\

\section{Further extending the class of simulatable operations}
\label{appendix:rational-projector-irrational-symplectic}
In this appendix, we will demonstrate that the class of symplectic operations simulatable using our method can be extended further than the rational symplectic matrices.
Specifically, there are certain instances whereby the projector ${1-SS^+}$, given in Eq.~(\ref{eq:appendix-projector}), is rational even when the symplectic matrix is irrational.

To understand why, consider that any symplectic matrix can be expressed as \cite{arvind1995}
\begin{align}
    \mqty(A&B\\C&D)=\mqty(A_0&0\\C_0&(A_0^{T})^{-1})\mqty(X&Y\\-Y&X)
\end{align}
where the second factor is an orthogonal symplectic matrix. Using $A=A_0X$ and $B=A_0Y$ we can write Eq.~(\ref{eq:sbardef}) as
\begin{align}
    \bar S=\mqty(X^TA_0^T\\Y^TA_0^T).
\end{align}
We can also express its pseudoinverse, given in Eq.~(\ref{eq:sbarpseudo}), as
\begin{align}
    \bar{S}^+=&(A_0XX^TA_0^T+A_0YY^TA_0^T)^{-1}\mqty(A_0X & A_0Y)\nonumber\\
    =&(A_0A_0^T)^{-1}\mqty(A_0X & A_0Y).
\end{align}
The rationality of the projection matrix ${1-SS^+}$ depends on the rationality of this matrix, which can be written as
\begin{align}
    \bar{S}\bar{S}^+
    =&\mqty(X^TA_0^T\\Y^TA_0^T)(A_0A_0^T)^{-1}\mqty(A_0X & A_0Y)\nonumber\\
    =&\mqty(X^TA_0^T\\Y^TA_0^T)(A_0^{-T}A_0^{-1})\mqty(A_0X & A_0Y)\nonumber\\
    =&\mqty(X^TX&X^TY\\Y^TX&Y^TY).
\end{align}
Therefore ${1-SS^+}$ will be rational as long as each of the blocks of this matrix are rational. I.e., the projector will be rational if all the elements of the matrices $X^TX,Y^TX$ are rational.

The projector can therefore in certain cases still be rational when the symplectic matrix is irrational. Namely, the matrix $A_0$ can be irrational while the projection matrix remains rational. 

There are also certain cases where the individual matrices $X,Y$ can be irrational while the projection matrix is rational. For example, consider the case that
\begin{align}
    X=\text{diag}(\text{cos}(\vec \theta)),\\
    Y=\text{diag}(\text{sin}(\vec \theta)).
\end{align}
We can rewrite these diagonal block matrices in terms of the tangent of the angles
\begin{align}
    (X^TX)_{jj}&=\text{cos}^2(\theta_j)=\frac{1}{\text{tan}^2( \theta_j)+1},\\
    (X^TY)_{jj}&=
     \text{cos}( \theta_j)\text{sin}( \theta_j)=
    \frac{\text{tan}( \theta_j)}{1+\text{tan}^2( \theta_j)},
\end{align}
from which we see that the projection matrix will be rational whenever $\text{tan}( \theta_j)\in\mathbb Q$ for all $j$.

Provided that the projection matrix in Eq.~(\ref{eq:appendix-projector}) is rational, it is possible to identify non-zero points of the PDF, by virtue of Appendix~\ref{sec:appendix-evaluation-l}. The constraint of rational symplectic matrices can therefore be relaxed. However, for simplicity, we choose to restrict to rational symplectic matrices in this work.

\section{Relationships between the classes of simulatable operations}
\label{appendix:relationships}
The class of operations which are shown to be efficiently simulatable in our work can be denoted by $\mathcal D$, which contains all operations deemed simulatable in Appendix \ref{appendix:rational-projector-irrational-symplectic}. For simplicity, throughout this work, we chose to denote the class of simulatable operations as those which belong to the class $\text{HW}(n)[\text{Sp}(2n,\mathbb Q)]$, i.e., those for which the symplectic matrix is rational.
\begin{figure*}[ht]
    \centering
    \includegraphics[width=0.8\linewidth]{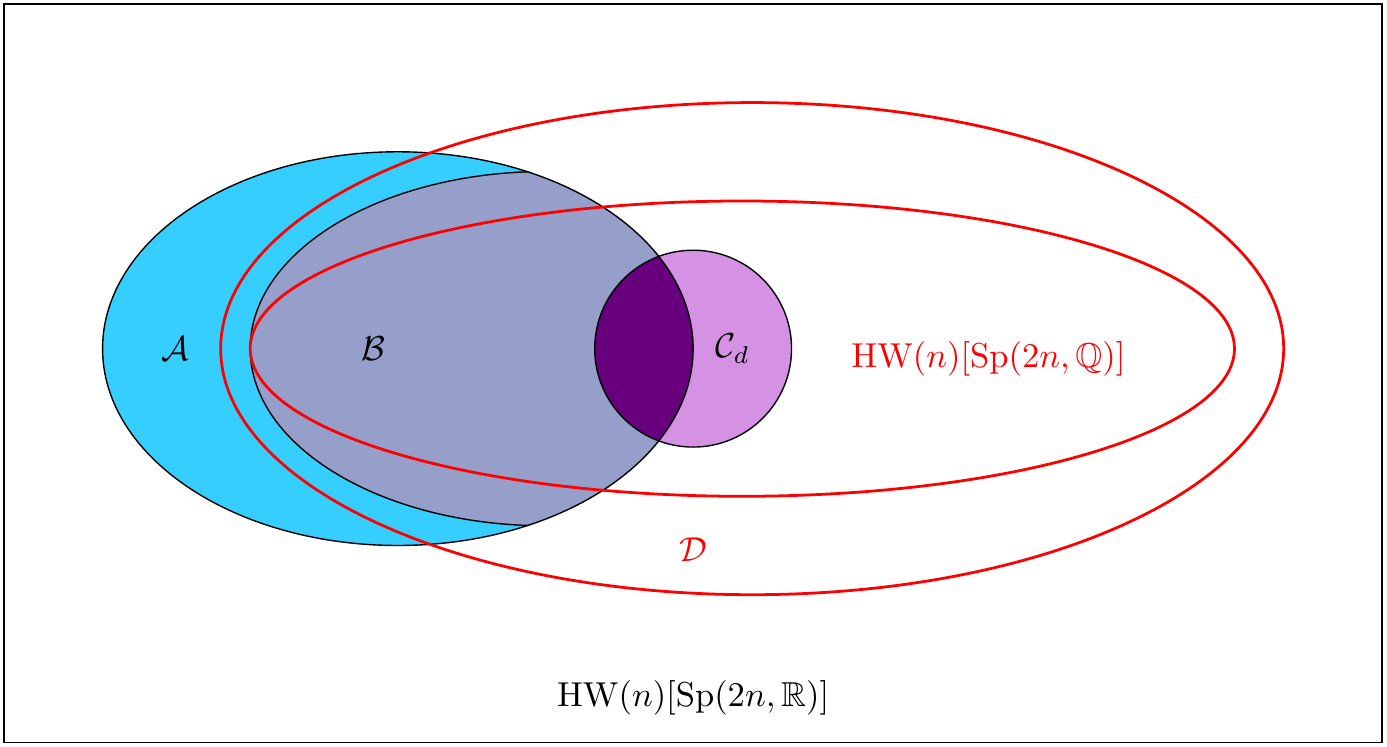}
    \caption{The classes of circuits considered in this work and previous works. 
    The class $\mathcal C_d$ refers to the Clifford group for dimension $d$. Classes $\mathcal A$ and $\mathcal B$ are defined in Ref.~\cite{calcluth2022} as the class of operations which are simulatable for single-mode and multi-mode measurement, respectively. $\mathcal A,\mathcal B,\mathcal C_d$ are all contained within the set of Gaussian operations $\text{HW}(n)[\text{Sp}(2n,\mathbb R)]$. The class $\text{HW}(n)[\text{Sp}(2n,\mathbb Q)]$ contains $\mathcal C_d$ but does not completely contain $\mathcal A$ nor $\mathcal B$. The class $\mathcal D$, as defined in this Appendix, contains $\mathcal B,\mathcal C_d$ but is not known to contain $\mathcal A$.  Note that the size of each of these regions in the diagram is arbitrary.
    }
    \label{fig:venn}
\end{figure*}

This class of operations $\text{HW}(n)[\text{Sp}(2n,\mathbb Q)]$ contains, in particular, all GKP Clifford operations for encoded qudits of any dimension, as was proven in Sec.~\ref{sec:clifford}.

We now recall and compare classes of operations that we demonstrated to be simulatable using different techniques in our previous work, Ref.~\cite{calcluth2022}, with those considered here. We previously demonstrated that circuits with input GKP states acted on by operations selected from a class $\mathcal B$ and measured in all modes with homodyne measurement are simulatable. This class was defined as
\begin{align}
\label{eq:classB}
    \mathcal B = \text{HW}(n)\times \text{DSp}(2n,\mathbb R)
\end{align}
where 
\begin{widetext}
    \begin{align}
    \label{eq:operations-dsp}
    \text{DSp}(2n,\mathbb R)=\left\{ \mqty(A_0&0\\C_0&(A_0^{T})^{-1})\mqty(\text{diag}(\cos\vec \theta)&\text{diag}(\sin\vec \theta)\\
    -\text{diag}(\sin\vec \theta)&\text{diag}(\cos\vec \theta)) : \det A_0 \ne 0, A_0^T = A_0,C_0^TA_0 = A_0^TC_0, \theta_j \in \Theta \right\}
    \end{align}
\end{widetext}
and
\begin{align}
    \label{eq:anglesset}
    \Theta=\{\theta \in \mathbb R: \cot\theta =u/v \in \mathbb Q_{(2)} \} \cup \{0,\pi\}.
\end{align}

The class $\mathcal B$ contains operations where the symplectic matrix can contain irrational elements, e.g. when $\theta_j = \pi/4$, despite satisfying the condition that $\cot\theta_j\in\mathbb Q_{(2)}$. This implies that $\mathcal B \not\subset \text{HW}(n)\times \text{Sp}(2n,\mathbb Q)$.

In Appendix \ref{appendix:rational-projector-irrational-symplectic} we demonstrated that it is possible to extend the class of simulatable operations beyond the group $\text{HW}(n)\times \text{Sp}(2n,\mathbb Q)$, to a larger set which we denote $\mathcal D$ and we show that $\mathcal B \subset \mathcal D$. This set contains all displacements $\text{HW}(n)$ and all symplectic matrices such that $X^TX$ and $X^TY$ are rational.

In our previous work \cite{calcluth2022} we demonstrated that is possible to simulate another class $\mathcal A$ which consists of symplectic operations whereby the top row of the symplectic matrix has a specific structure. That matrix does not necessarily satisfy any constraints in the other elements and so we cannot conclude that $\mathcal D$ nor $\text{HW}(n)\times \text{Sp}(2n,\mathbb Q)$ contains $\mathcal A$.

We have included a figure, Fig.~\ref{fig:venn}, to show the containment of each of these classes of operations, with respect to the previous classes identified in Ref.~\cite{calcluth2022}.

\section{Adaptive circuits with modular homodyne measurements}
\label{sec:appendix-adaptive-mod}
Although not required for the results of this paper, we provide an additional observation in this appendix. We demonstrate how to efficiently sample from a circuit that makes use of modular measurements. This method involves producing random integers selected from a finite set of integers.
 
Quantum circuits involving GKP states often make use of modular homodyne measurements. These are measurements in position or momentum modulo some period. Formally we define some period $T/\sqrt\pi\in \mathbb Q$ such that the recorded measurement result in position or momentum, $x_1$, is recorded as $x_1\mod T$. For example, Pauli $\hat Z$ measurements in the GKP framework~\cite{gottesman2001} are measurements in position modulo $T=2\sqrt\pi$. If the measurement result $x_1$ is closest to $x_1\mod 2\sqrt\pi=0$, the measurement corresponds to a measurement of the logical qubit state $\ket 0$. If the measurement result $x_1$ is closest to $x_1\mod 2\sqrt\pi=\sqrt\pi$, the measurement corresponds to a measurement of the logical qubit state $\ket 1$.

An adaptive circuit with feed-forward operations that makes use of modular measurements will use the value of $x_1 \mod 2\sqrt\pi$ to determine future operations. In the case of Pauli $\hat Z$ measurements, we can define two possible operations which could be performed on the remaining modes, depending on which outcome is measured.

The PDF of a unitary non-adaptive operation $U_0$ followed by a measurement of mode $1$ can be represented as
\begin{align}
    \text{PDF}(x_1)=&\sum_{\vec m\in \mathbb Z^n}\delta(x_1-\sqrt\pi \left((R^T)^{-1}(\vec t+2\vec m)\right)_1-c_1)
\end{align}
which is equivalent to identifying that the possible measurement values of $x_1$ can be given by
\begin{align}
    x_1=\sqrt\pi \left(R^{-T}(\vec t+2\vec m)\right)_1 + c_1 \quad \forall \quad  \vec m \in \mathbb Z^n.
\end{align}

To identify the possible outcomes of $x_1\mod \sqrt\pi k$, we calculate
\begin{align}
    x_1\mod k\sqrt\pi =\sqrt\pi \left(R^{-T}(\vec t+2\vec m)\right)_1 + c_1 \mod k\sqrt\pi.
\end{align}
The matrix $R$ is rational and so we can write~\cite{calcluth2022}
\begin{align}
    ((R^{-T})2\vec m)_1=\frac{u}{v} m^*
\end{align}
which reduces the random vector of integers to a single integer $m^*\in \mathbb Z$, and a period $\frac{u}{v}\in\mathbb Q$, which depends on the first row of the matrix $R^{-T}$.

This allows us to simplify the possible measurement outcomes to
\begin{align}
    \bar x_1 =& x_1\mod k\sqrt\pi \nonumber \\
    =&\sqrt\pi \left(R^{-T}\vec t\right)_1+\sqrt\pi \frac{u}{v}m^* + c_1 \mod k\sqrt\pi
\end{align}
where we can restrict to at most $vk$ possible outcomes parameterized by $\bar m\in\{0,1,2,vk-1\}$, which each occur with equal probability. Simulation of measurement consists of choosing a random value of $\bar m$ from the finite set of possible integers.

\begin{widetext}
Following the adaptive routine, we then choose a new operator $U_1(x_1)$ dependent on the measured value of $x_1$ and simulate the circuit $U_1(x_1)U_0$. This will provide us with points of the form
\begin{align}
    \text{PDF}(\vec x)=\sum_{\vec m} \delta(x_1-\left(\sqrt\pi R'^{-T}(\vec t'+2\vec m')+\vec c'\right)_1)\dots \delta(x_n-\left(\sqrt\pi R'^{-T}(\vec t'+2\vec m')+\vec c'\right)_n).
\end{align}
Choosing $x_1=\bar x_1$ and assuming no operations have been applied to the measured mode we have
\begin{align}
    \text{PDF}(\vec x)=\sum_{\vec m} \delta( \frac{u}{v}\bar m-2\left(R'^{-T}\vec m'\right)_1)\dots \delta(x_n-\left(\sqrt\pi R'^{-T}(\vec t'+2\vec m')+\vec c'\right)_n).
\end{align}
This expression can be simplified to a summation over $n-1$ integers.
\end{widetext}

\bibliographystyle{apsrev4-2}
\bibliography{./main.bib}

\begin{thebibliography}{56}%
\makeatletter
\providecommand \@ifxundefined [1]{%
 \@ifx{#1\undefined}
}%
\providecommand \@ifnum [1]{%
 \ifnum #1\expandafter \@firstoftwo
 \else \expandafter \@secondoftwo
 \fi
}%
\providecommand \@ifx [1]{%
 \ifx #1\expandafter \@firstoftwo
 \else \expandafter \@secondoftwo
 \fi
}%
\providecommand \natexlab [1]{#1}%
\providecommand \enquote  [1]{``#1''}%
\providecommand \bibnamefont  [1]{#1}%
\providecommand \bibfnamefont [1]{#1}%
\providecommand \citenamefont [1]{#1}%
\providecommand \href@noop [0]{\@secondoftwo}%
\providecommand \href [0]{\begingroup \@sanitize@url \@href}%
\providecommand \@href[1]{\@@startlink{#1}\@@href}%
\providecommand \@@href[1]{\endgroup#1\@@endlink}%
\providecommand \@sanitize@url [0]{\catcode `\\12\catcode `\$12\catcode
  `\&12\catcode `\#12\catcode `\^12\catcode `\_12\catcode `\%12\relax}%
\providecommand \@@startlink[1]{}%
\providecommand \@@endlink[0]{}%
\providecommand \url  [0]{\begingroup\@sanitize@url \@url }%
\providecommand \@url [1]{\endgroup\@href {#1}{\urlprefix }}%
\providecommand \urlprefix  [0]{URL }%
\providecommand \Eprint [0]{\href }%
\providecommand \doibase [0]{https://doi.org/}%
\providecommand \selectlanguage [0]{\@gobble}%
\providecommand \bibinfo  [0]{\@secondoftwo}%
\providecommand \bibfield  [0]{\@secondoftwo}%
\providecommand \translation [1]{[#1]}%
\providecommand \BibitemOpen [0]{}%
\providecommand \bibitemStop [0]{}%
\providecommand \bibitemNoStop [0]{.\EOS\space}%
\providecommand \EOS [0]{\spacefactor3000\relax}%
\providecommand \BibitemShut  [1]{\csname bibitem#1\endcsname}%
\let\auto@bib@innerbib\@empty
\bibitem [{\citenamefont {Chitambar}\ and\ \citenamefont
  {Gour}(2019)}]{Gour-2019}%
  \BibitemOpen
  \bibfield  {author} {\bibinfo {author} {\bibfnamefont {E.}~\bibnamefont
  {Chitambar}}\ and\ \bibinfo {author} {\bibfnamefont {G.}~\bibnamefont
  {Gour}},\ }\href {https://doi.org/10.1103/RevModPhys.91.025001} {\bibfield
  {journal} {\bibinfo  {journal} {Rev. Mod. Phys.}\ }\textbf {\bibinfo {volume}
  {91}},\ \bibinfo {pages} {025001} (\bibinfo {year} {2019})}\BibitemShut
  {NoStop}%
\bibitem [{\citenamefont {Gottesman}(1997)}]{gottesman1997}%
  \BibitemOpen
  \bibfield  {author} {\bibinfo {author} {\bibfnamefont {D.}~\bibnamefont
  {Gottesman}},\ }\href {https://arxiv.org/abs/quant-ph/9705052v1} {\bibfield
  {journal} {\bibinfo  {journal} {PhD Thesis}\ } (\bibinfo {year} {1997})},\
  \bibinfo {note} {arXiv:quant-ph/9705052v1}\BibitemShut {NoStop}%
\bibitem [{\citenamefont {Gottesman}(1999)}]{gottesman1999}%
  \BibitemOpen
  \bibfield  {author} {\bibinfo {author} {\bibfnamefont {D.}~\bibnamefont
  {Gottesman}},\ }\href {https://arxiv.org/abs/quant-ph/9807006v1} {\emph
  {\bibinfo {title} {The Heisenberg representation of quantum computers}}},\
  edited by\ \bibinfo {editor} {\bibfnamefont {S.~P.}\ \bibnamefont {Corney}},
  \bibinfo {editor} {\bibfnamefont {R.}~\bibnamefont {Delbourgo}},\ and\
  \bibinfo {editor} {\bibfnamefont {P.~D.}\ \bibnamefont {Jarvis}},\ Group22:
  Proceedings of the XXII International Colloquium on Group Theoretical Methods
  in Physics\ (\bibinfo  {publisher} {Cambridge, MA, International Press},\
  \bibinfo {year} {1999})\ pp.\ \bibinfo {pages} {32--43},\ \bibinfo {note}
  {arXiv:quant-ph/9807006}\BibitemShut {NoStop}%
\bibitem [{\citenamefont {Nielsen}\ and\ \citenamefont
  {Chuang}(2000)}]{nielsen2000}%
  \BibitemOpen
  \bibfield  {author} {\bibinfo {author} {\bibfnamefont {M.~A.}\ \bibnamefont
  {Nielsen}}\ and\ \bibinfo {author} {\bibfnamefont {I.~L.}\ \bibnamefont
  {Chuang}},\ }\href@noop {} {\emph {\bibinfo {title} {{Quantum Computation and
  Quantum Information}}}}\ (\bibinfo  {publisher} {Cambridge University
  Press},\ \bibinfo {year} {2000})\BibitemShut {NoStop}%
\bibitem [{\citenamefont {Bravyi}\ and\ \citenamefont
  {Kitaev}(2005)}]{bravyi2005}%
  \BibitemOpen
  \bibfield  {author} {\bibinfo {author} {\bibfnamefont {S.}~\bibnamefont
  {Bravyi}}\ and\ \bibinfo {author} {\bibfnamefont {A.}~\bibnamefont
  {Kitaev}},\ }\href {https://doi.org/10.1103/PhysRevA.71.022316} {\bibfield
  {journal} {\bibinfo  {journal} {Physical Review A}\ }\textbf {\bibinfo
  {volume} {71}},\ \bibinfo {pages} {022316} (\bibinfo {year}
  {2005})}\BibitemShut {NoStop}%
\bibitem [{\citenamefont {Reichardt}(2005)}]{reichardt2005}%
  \BibitemOpen
  \bibfield  {author} {\bibinfo {author} {\bibfnamefont {B.~W.}\ \bibnamefont
  {Reichardt}},\ }\href {https://doi.org/10.1007/s11128-005-7654-8} {\bibfield
  {journal} {\bibinfo  {journal} {Quantum Information Processing}\ }\textbf
  {\bibinfo {volume} {4}},\ \bibinfo {pages} {251} (\bibinfo {year}
  {2005})}\BibitemShut {NoStop}%
\bibitem [{\citenamefont {Bartlett}\ \emph {et~al.}(2002)\citenamefont
  {Bartlett}, \citenamefont {Sanders}, \citenamefont {Braunstein},\ and\
  \citenamefont {Nemoto}}]{bartlett2002}%
  \BibitemOpen
  \bibfield  {author} {\bibinfo {author} {\bibfnamefont {S.~D.}\ \bibnamefont
  {Bartlett}}, \bibinfo {author} {\bibfnamefont {B.~C.}\ \bibnamefont
  {Sanders}}, \bibinfo {author} {\bibfnamefont {S.~L.}\ \bibnamefont
  {Braunstein}},\ and\ \bibinfo {author} {\bibfnamefont {K.}~\bibnamefont
  {Nemoto}},\ }\href {https://doi.org/10.1103/PhysRevLett.88.097904} {\bibfield
   {journal} {\bibinfo  {journal} {Phys. Rev. Lett.}\ }\textbf {\bibinfo
  {volume} {88}},\ \bibinfo {pages} {097904} (\bibinfo {year}
  {2002})}\BibitemShut {NoStop}%
\bibitem [{\citenamefont {Mari}\ and\ \citenamefont {Eisert}(2012)}]{mari2012}%
  \BibitemOpen
  \bibfield  {author} {\bibinfo {author} {\bibfnamefont {A.}~\bibnamefont
  {Mari}}\ and\ \bibinfo {author} {\bibfnamefont {J.}~\bibnamefont {Eisert}},\
  }\href {https://doi.org/10.1103/PhysRevLett.109.230503} {\bibfield  {journal}
  {\bibinfo  {journal} {Physical Review Letters}\ }\textbf {\bibinfo {volume}
  {109}},\ \bibinfo {pages} {230503} (\bibinfo {year} {2012})}\BibitemShut
  {NoStop}%
\bibitem [{\citenamefont {Veitch}\ \emph {et~al.}(2012)\citenamefont {Veitch},
  \citenamefont {Ferrie}, \citenamefont {Gross},\ and\ \citenamefont
  {Emerson}}]{veitch2012}%
  \BibitemOpen
  \bibfield  {author} {\bibinfo {author} {\bibfnamefont {V.}~\bibnamefont
  {Veitch}}, \bibinfo {author} {\bibfnamefont {C.}~\bibnamefont {Ferrie}},
  \bibinfo {author} {\bibfnamefont {D.}~\bibnamefont {Gross}},\ and\ \bibinfo
  {author} {\bibfnamefont {J.}~\bibnamefont {Emerson}},\ }\href
  {https://doi.org/10.1088/1367-2630/14/11/113011} {\bibfield  {journal}
  {\bibinfo  {journal} {New Journal of Physics}\ }\textbf {\bibinfo {volume}
  {14}},\ \bibinfo {pages} {113011} (\bibinfo {year} {2012})}\BibitemShut
  {NoStop}%
\bibitem [{\citenamefont {Albarelli}\ \emph {et~al.}(2018)\citenamefont
  {Albarelli}, \citenamefont {Genoni}, \citenamefont {Paris},\ and\
  \citenamefont {Ferraro}}]{albarelli2018}%
  \BibitemOpen
  \bibfield  {author} {\bibinfo {author} {\bibfnamefont {F.}~\bibnamefont
  {Albarelli}}, \bibinfo {author} {\bibfnamefont {M.~G.}\ \bibnamefont
  {Genoni}}, \bibinfo {author} {\bibfnamefont {M.~G.~A.}\ \bibnamefont
  {Paris}},\ and\ \bibinfo {author} {\bibfnamefont {A.}~\bibnamefont
  {Ferraro}},\ }\href {https://doi.org/10.1103/PhysRevA.98.052350} {\bibfield
  {journal} {\bibinfo  {journal} {Physical Review A}\ }\textbf {\bibinfo
  {volume} {98}},\ \bibinfo {pages} {052350} (\bibinfo {year}
  {2018})}\BibitemShut {NoStop}%
\bibitem [{\citenamefont {Takagi}\ and\ \citenamefont
  {Zhuang}(2018)}]{takagi2018}%
  \BibitemOpen
  \bibfield  {author} {\bibinfo {author} {\bibfnamefont {R.}~\bibnamefont
  {Takagi}}\ and\ \bibinfo {author} {\bibfnamefont {Q.}~\bibnamefont
  {Zhuang}},\ }\href {https://doi.org/10.1103/PhysRevA.97.062337} {\bibfield
  {journal} {\bibinfo  {journal} {Physical Review A}\ }\textbf {\bibinfo
  {volume} {97}},\ \bibinfo {pages} {062337} (\bibinfo {year}
  {2018})}\BibitemShut {NoStop}%
\bibitem [{\citenamefont {Gottesman}\ \emph {et~al.}(2001)\citenamefont
  {Gottesman}, \citenamefont {Kitaev},\ and\ \citenamefont
  {Preskill}}]{gottesman2001}%
  \BibitemOpen
  \bibfield  {author} {\bibinfo {author} {\bibfnamefont {D.}~\bibnamefont
  {Gottesman}}, \bibinfo {author} {\bibfnamefont {A.}~\bibnamefont {Kitaev}},\
  and\ \bibinfo {author} {\bibfnamefont {J.}~\bibnamefont {Preskill}},\ }\href
  {https://doi.org/10.1103/PhysRevA.64.012310} {\bibfield  {journal} {\bibinfo
  {journal} {Physical Review A}\ }\textbf {\bibinfo {volume} {64}},\ \bibinfo
  {pages} {012310} (\bibinfo {year} {2001})}\BibitemShut {NoStop}%
\bibitem [{\citenamefont {Lloyd}\ and\ \citenamefont
  {Braunstein}(1999)}]{lloyd1999}%
  \BibitemOpen
  \bibfield  {author} {\bibinfo {author} {\bibfnamefont {S.}~\bibnamefont
  {Lloyd}}\ and\ \bibinfo {author} {\bibfnamefont {S.~L.}\ \bibnamefont
  {Braunstein}},\ }\href {https://doi.org/10.1103/PhysRevLett.82.1784}
  {\bibfield  {journal} {\bibinfo  {journal} {Phys. Rev. Lett.}\ }\textbf
  {\bibinfo {volume} {82}},\ \bibinfo {pages} {1784} (\bibinfo {year}
  {1999})}\BibitemShut {NoStop}%
\bibitem [{\citenamefont {Baragiola}\ \emph {et~al.}(2019)\citenamefont
  {Baragiola}, \citenamefont {Pantaleoni}, \citenamefont {Alexander},
  \citenamefont {Karanjai},\ and\ \citenamefont {Menicucci}}]{baragiola2019}%
  \BibitemOpen
  \bibfield  {author} {\bibinfo {author} {\bibfnamefont {B.~Q.}\ \bibnamefont
  {Baragiola}}, \bibinfo {author} {\bibfnamefont {G.}~\bibnamefont
  {Pantaleoni}}, \bibinfo {author} {\bibfnamefont {R.~N.}\ \bibnamefont
  {Alexander}}, \bibinfo {author} {\bibfnamefont {A.}~\bibnamefont
  {Karanjai}},\ and\ \bibinfo {author} {\bibfnamefont {N.~C.}\ \bibnamefont
  {Menicucci}},\ }\href {https://doi.org/10.1103/PhysRevLett.123.200502}
  {\bibfield  {journal} {\bibinfo  {journal} {Phys. Rev. Lett.}\ }\textbf
  {\bibinfo {volume} {123}},\ \bibinfo {pages} {200502} (\bibinfo {year}
  {2019})}\BibitemShut {NoStop}%
\bibitem [{\citenamefont {Yamasaki}\ \emph {et~al.}(2020)\citenamefont
  {Yamasaki}, \citenamefont {Matsuura},\ and\ \citenamefont
  {Koashi}}]{yamasaki2020}%
  \BibitemOpen
  \bibfield  {author} {\bibinfo {author} {\bibfnamefont {H.}~\bibnamefont
  {Yamasaki}}, \bibinfo {author} {\bibfnamefont {T.}~\bibnamefont {Matsuura}},\
  and\ \bibinfo {author} {\bibfnamefont {M.}~\bibnamefont {Koashi}},\ }\href
  {https://doi.org/10.1103/PhysRevResearch.2.023270} {\bibfield  {journal}
  {\bibinfo  {journal} {Physical Review Research}\ }\textbf {\bibinfo {volume}
  {2}},\ \bibinfo {pages} {023270} (\bibinfo {year} {2020})}\BibitemShut
  {NoStop}%
\bibitem [{\citenamefont {de~Beaudrap}(2013)}]{beaudrap2013}%
  \BibitemOpen
  \bibfield  {author} {\bibinfo {author} {\bibfnamefont {N.}~\bibnamefont
  {de~Beaudrap}},\ }\href {https://arxiv.org/abs/1102.3354} {\bibfield
  {journal} {\bibinfo  {journal} {Quantum Information \& Computation}\ }\textbf
  {\bibinfo {volume} {13}},\ \bibinfo {pages} {73} (\bibinfo {year} {2013})},\
  \bibinfo {note} {arXiv:1102.3354}\BibitemShut {NoStop}%
\bibitem [{\citenamefont {Gheorghiu}(2014)}]{gheorghiu2014}%
  \BibitemOpen
  \bibfield  {author} {\bibinfo {author} {\bibfnamefont {V.}~\bibnamefont
  {Gheorghiu}},\ }\href
  {https://doi.org/https://doi.org/10.1016/j.physleta.2013.12.009} {\bibfield
  {journal} {\bibinfo  {journal} {Physics Letters A}\ }\textbf {\bibinfo
  {volume} {378}},\ \bibinfo {pages} {505 } (\bibinfo {year}
  {2014})}\BibitemShut {NoStop}%
\bibitem [{Note1()}]{Note1}%
  \BibitemOpen
  \bibinfo {note} {Note that in the main text, we simplify the class of
  simulatable operations to those which have a rational symplectic matrix.
  However, the class of simulatable operations also includes those given in the
  multimode case of Ref.~\cite {calcluth2022}. We provide the broader
  requirements of the class of simulatable symplectic matrices in Appendix~\ref
  {appendix:rational-projector-irrational-symplectic}}\BibitemShut {NoStop}%
\bibitem [{\citenamefont {Garc\'{\i}a-\'Alvarez}\ \emph
  {et~al.}(2020)\citenamefont {Garc\'{\i}a-\'Alvarez}, \citenamefont
  {Calcluth}, \citenamefont {Ferraro},\ and\ \citenamefont
  {Ferrini}}]{garcia-alvarez2020}%
  \BibitemOpen
  \bibfield  {author} {\bibinfo {author} {\bibfnamefont {L.}~\bibnamefont
  {Garc\'{\i}a-\'Alvarez}}, \bibinfo {author} {\bibfnamefont {C.}~\bibnamefont
  {Calcluth}}, \bibinfo {author} {\bibfnamefont {A.}~\bibnamefont {Ferraro}},\
  and\ \bibinfo {author} {\bibfnamefont {G.}~\bibnamefont {Ferrini}},\ }\href
  {https://doi.org/10.1103/PhysRevResearch.2.043322} {\bibfield  {journal}
  {\bibinfo  {journal} {Phys. Rev. Research}\ }\textbf {\bibinfo {volume}
  {2}},\ \bibinfo {pages} {043322} (\bibinfo {year} {2020})}\BibitemShut
  {NoStop}%
\bibitem [{\citenamefont {Calcluth}\ \emph {et~al.}(2022)\citenamefont
  {Calcluth}, \citenamefont {Ferraro},\ and\ \citenamefont
  {Ferrini}}]{calcluth2022}%
  \BibitemOpen
  \bibfield  {author} {\bibinfo {author} {\bibfnamefont {C.}~\bibnamefont
  {Calcluth}}, \bibinfo {author} {\bibfnamefont {A.}~\bibnamefont {Ferraro}},\
  and\ \bibinfo {author} {\bibfnamefont {G.}~\bibnamefont {Ferrini}},\ }\href
  {https://arxiv.org/abs/2203.11182} {\bibfield  {journal} {\bibinfo  {journal}
  {arXiv:2203.11182}\ } (\bibinfo {year} {2022})}\BibitemShut {NoStop}%
\bibitem [{\citenamefont {Garc{\'i}a-{\'A}lvarez}\ \emph
  {et~al.}(2021)\citenamefont {Garc{\'i}a-{\'A}lvarez}, \citenamefont
  {Ferraro},\ and\ \citenamefont {Ferrini}}]{garcia-alvarez2019}%
  \BibitemOpen
  \bibfield  {author} {\bibinfo {author} {\bibfnamefont {L.}~\bibnamefont
  {Garc{\'i}a-{\'A}lvarez}}, \bibinfo {author} {\bibfnamefont {A.}~\bibnamefont
  {Ferraro}},\ and\ \bibinfo {author} {\bibfnamefont {G.}~\bibnamefont
  {Ferrini}},\ }in\ \href {https://doi.org/10.1007/978-981-15-5191-8_9} {\emph
  {\bibinfo {booktitle} {International Symposium on Mathematics, Quantum
  Theory, and Cryptography}}},\ \bibinfo {editor} {edited by\ \bibinfo {editor}
  {\bibfnamefont {T.}~\bibnamefont {Takagi}}, \bibinfo {editor} {\bibfnamefont
  {M.}~\bibnamefont {Wakayama}}, \bibinfo {editor} {\bibfnamefont
  {K.}~\bibnamefont {Tanaka}}, \bibinfo {editor} {\bibfnamefont
  {N.}~\bibnamefont {Kunihiro}}, \bibinfo {editor} {\bibfnamefont
  {K.}~\bibnamefont {Kimoto}},\ and\ \bibinfo {editor} {\bibfnamefont
  {Y.}~\bibnamefont {Ikematsu}}}\ (\bibinfo  {publisher} {Springer Singapore},\
  \bibinfo {address} {Singapore},\ \bibinfo {year} {2021})\ pp.\ \bibinfo
  {pages} {79--92}\BibitemShut {NoStop}%
\bibitem [{\citenamefont {Rahimi-Keshari}\ \emph {et~al.}(2016)\citenamefont
  {Rahimi-Keshari}, \citenamefont {Ralph},\ and\ \citenamefont
  {Caves}}]{rahimi-keshari2016}%
  \BibitemOpen
  \bibfield  {author} {\bibinfo {author} {\bibfnamefont {S.}~\bibnamefont
  {Rahimi-Keshari}}, \bibinfo {author} {\bibfnamefont {T.~C.}\ \bibnamefont
  {Ralph}},\ and\ \bibinfo {author} {\bibfnamefont {C.~M.}\ \bibnamefont
  {Caves}},\ }\href {https://doi.org/10.1103/PhysRevX.6.021039} {\bibfield
  {journal} {\bibinfo  {journal} {Physical Review X}\ }\textbf {\bibinfo
  {volume} {6}},\ \bibinfo {pages} {021039} (\bibinfo {year}
  {2016})}\BibitemShut {NoStop}%
\bibitem [{Note2()}]{Note2}%
  \BibitemOpen
  \bibinfo {note} {The Heisenberg-Weyl group $\protect \text {HW}(n)$ is a
  normal subgroup of the semi-direct product of $\protect \text {HW}(n)$ and
  $\protect \text {Sp}(2n,\protect \mathbb Q)$, which we indicate by $\protect
  \text {HW}(n)[\protect \text {Sp}(2n,\protect \mathbb Q)]$. Indeed, the
  subgroup $\protect \text {HW}(n)$ is invariant under conjugation by any
  element of $\protect \text {HW}(n)[\protect \text {Sp}(2n,\protect \mathbb
  Q)]$. Therefore, the full group of simulatable operations is specified by the
  semi-direct product of these two subgroups~\cite {dummit1991}.}\BibitemShut
  {Stop}%
\bibitem [{\citenamefont {Ferraro}\ \emph {et~al.}(2005)\citenamefont
  {Ferraro}, \citenamefont {Olivares},\ and\ \citenamefont
  {Paris}}]{ferraro2005}%
  \BibitemOpen
  \bibfield  {author} {\bibinfo {author} {\bibfnamefont {A.}~\bibnamefont
  {Ferraro}}, \bibinfo {author} {\bibfnamefont {S.}~\bibnamefont {Olivares}},\
  and\ \bibinfo {author} {\bibfnamefont {M.~G.~A.}\ \bibnamefont {Paris}},\
  }\href@noop {} {\emph {\bibinfo {title} {Gaussian States in Quantum
  Information}}}\ (\bibinfo  {publisher} {Bibliopolis},\ \bibinfo {address}
  {Napoli},\ \bibinfo {year} {2005})\ \bibinfo {note}
  {\href{https://arxiv.org/abs/quant-ph/0503237}{arXiv:quant-ph/0503237}}\BibitemShut
  {NoStop}%
\bibitem [{\citenamefont {Serafini}(2017)}]{serafini2017}%
  \BibitemOpen
  \bibfield  {author} {\bibinfo {author} {\bibfnamefont {A.}~\bibnamefont
  {Serafini}},\ }\href@noop {} {\emph {\bibinfo {title} {{Quantum continuous
  variables : a primer of theoretical methods}}}}\ (\bibinfo  {publisher} {CRC
  Press, Taylor {\&} Francis Group},\ \bibinfo {address} {Boca Raton, FL},\
  \bibinfo {year} {2017})\BibitemShut {NoStop}%
\bibitem [{\citenamefont {Kok}\ and\ \citenamefont {Lovett}(2010)}]{kok2010}%
  \BibitemOpen
  \bibfield  {author} {\bibinfo {author} {\bibfnamefont {P.}~\bibnamefont
  {Kok}}\ and\ \bibinfo {author} {\bibfnamefont {B.~W.}\ \bibnamefont
  {Lovett}},\ }\href@noop {} {\emph {\bibinfo {title} {Introduction to optical
  quantum information processing}}}\ (\bibinfo  {publisher} {Cambridge
  university press},\ \bibinfo {year} {2010})\BibitemShut {NoStop}%
\bibitem [{\citenamefont {Sakurai}\ and\ \citenamefont
  {Napolitano}(2017)}]{sakurai}%
  \BibitemOpen
  \bibfield  {author} {\bibinfo {author} {\bibfnamefont {J.~J.}\ \bibnamefont
  {Sakurai}}\ and\ \bibinfo {author} {\bibfnamefont {J.}~\bibnamefont
  {Napolitano}},\ }\href {https://doi.org/10.1017/9781108499996} {\emph
  {\bibinfo {title} {Modern Quantum Mechanics}}},\ \bibinfo {edition} {2nd}\
  ed.\ (\bibinfo  {publisher} {Cambridge University Press},\ \bibinfo {year}
  {2017})\BibitemShut {NoStop}%
\bibitem [{\citenamefont {Gerry}\ \emph {et~al.}(2005)\citenamefont {Gerry},
  \citenamefont {Knight},\ and\ \citenamefont {Knight}}]{gerry2005}%
  \BibitemOpen
  \bibfield  {author} {\bibinfo {author} {\bibfnamefont {C.}~\bibnamefont
  {Gerry}}, \bibinfo {author} {\bibfnamefont {P.}~\bibnamefont {Knight}},\ and\
  \bibinfo {author} {\bibfnamefont {P.~L.}\ \bibnamefont {Knight}},\
  }\href@noop {} {\emph {\bibinfo {title} {Introductory quantum optics}}}\
  (\bibinfo  {publisher} {Cambridge university press},\ \bibinfo {year}
  {2005})\BibitemShut {NoStop}%
\bibitem [{\citenamefont {Moore}(1920)}]{moore1920}%
  \BibitemOpen
  \bibfield  {author} {\bibinfo {author} {\bibfnamefont {E.~H.}\ \bibnamefont
  {Moore}},\ }\href@noop {} {\bibfield  {journal} {\bibinfo  {journal} {Bull.
  Am. Math. Soc.}\ }\textbf {\bibinfo {volume} {26}},\ \bibinfo {pages} {394}
  (\bibinfo {year} {1920})}\BibitemShut {NoStop}%
\bibitem [{\citenamefont {Penrose}(1955)}]{penrose1955}%
  \BibitemOpen
  \bibfield  {author} {\bibinfo {author} {\bibfnamefont {R.}~\bibnamefont
  {Penrose}},\ }in\ \href@noop {} {\emph {\bibinfo {booktitle} {Mathematical
  proceedings of the Cambridge philosophical society}}},\ Vol.~\bibinfo
  {volume} {51}\ (\bibinfo {organization} {Cambridge University Press},\
  \bibinfo {year} {1955})\ pp.\ \bibinfo {pages} {406--413}\BibitemShut
  {NoStop}%
\bibitem [{\citenamefont {Ben-Israel}\ and\ \citenamefont
  {Greville}(2003)}]{ben2003}%
  \BibitemOpen
  \bibfield  {author} {\bibinfo {author} {\bibfnamefont {A.}~\bibnamefont
  {Ben-Israel}}\ and\ \bibinfo {author} {\bibfnamefont {T.~N.}\ \bibnamefont
  {Greville}},\ }\href {https://doi.org/10.1007/b97366} {\emph {\bibinfo
  {title} {Generalized inverses: theory and applications}}},\ Vol.~\bibinfo
  {volume} {15}\ (\bibinfo  {publisher} {Springer Science \& Business Media},\
  \bibinfo {year} {2003})\BibitemShut {NoStop}%
\bibitem [{\citenamefont {Newman}(1972)}]{newman1972}%
  \BibitemOpen
  \bibfield  {author} {\bibinfo {author} {\bibfnamefont {M.}~\bibnamefont
  {Newman}},\ }\href@noop {} {\emph {\bibinfo {title} {Integral Matrices}}},\
  \bibinfo {series} {Pure and Applied Mathematics; a Series of Monographs and
  Textbooks}\ No.\ \bibinfo {number} {v. 45}\ (\bibinfo  {publisher} {Academic
  Press},\ \bibinfo {year} {1972})\BibitemShut {NoStop}%
\bibitem [{\citenamefont {Newman}(1997)}]{newman1997}%
  \BibitemOpen
  \bibfield  {author} {\bibinfo {author} {\bibfnamefont {M.}~\bibnamefont
  {Newman}},\ }\href@noop {} {\bibfield  {journal} {\bibinfo  {journal} {Linear
  algebra and its applications}\ }\textbf {\bibinfo {volume} {254}},\ \bibinfo
  {pages} {367} (\bibinfo {year} {1997})}\BibitemShut {NoStop}%
\bibitem [{sta(2022{\natexlab{a}})}]{stackexchangesmith}%
  \BibitemOpen
  \href@noop {} {\bibinfo {title} {Integer eigenvectors of a rational matrix}}
  (\bibinfo {year} {2022}{\natexlab{a}}),\ \bibinfo {note} {mathematics Stack
  Exchange. Available at
  \url{https://math.stackexchange.com/questions/4391454/integer-eigenvectors-of-a-rational-matrix/4391951}
  (accessed: 2022-04-26)}\BibitemShut {NoStop}%
\bibitem [{Note3()}]{Note3}%
  \BibitemOpen
  \bibinfo {note} {Alternative previous results~\cite
  {Juani-thesis,bermejo-vega2016} also exist for the simulation of CV circuits
  in the form of normalizer circuits. These results provide a numerical method
  to simulate non-adaptive normalizer circuits in the weak sense~\cite
  {jozsa2014}, i.e. it is possible to sample the output of a non-adaptive
  circuit. However, adaptivity is required for magic state distillation and so
  these results alone do not allow us to conclude that the vacuum is
  responsible for providing quantum advantage.}\BibitemShut {Stop}%
\bibitem [{\citenamefont {Jozsa}\ and\ \citenamefont {Van
  Den~Nest}(2014)}]{jozsa2014}%
  \BibitemOpen
  \bibfield  {author} {\bibinfo {author} {\bibfnamefont {R.}~\bibnamefont
  {Jozsa}}\ and\ \bibinfo {author} {\bibfnamefont {M.}~\bibnamefont {Van
  Den~Nest}},\ }\href@noop {} {\bibfield  {journal} {\bibinfo  {journal}
  {Quantum Information \& Computation}\ }\textbf {\bibinfo {volume} {14}},\
  \bibinfo {pages} {633} (\bibinfo {year} {2014})}\BibitemShut {NoStop}%
\bibitem [{\citenamefont {Arora}\ and\ \citenamefont
  {Barak}(2009)}]{arora2009}%
  \BibitemOpen
  \bibfield  {author} {\bibinfo {author} {\bibfnamefont {S.}~\bibnamefont
  {Arora}}\ and\ \bibinfo {author} {\bibfnamefont {B.}~\bibnamefont {Barak}},\
  }\href {https://doi.org/10.1017/CBO9780511804090} {\emph {\bibinfo {title}
  {Computational {Complexity}: {A} {Modern} {Approach}}}}\ (\bibinfo
  {publisher} {Cambridge University Press},\ \bibinfo {address} {Cambridge},\
  \bibinfo {year} {2009})\BibitemShut {NoStop}%
\bibitem [{\citenamefont {Mollin}(2008)}]{mollin2008}%
  \BibitemOpen
  \bibfield  {author} {\bibinfo {author} {\bibfnamefont {R.~A.}\ \bibnamefont
  {Mollin}},\ }\href {https://doi.org/10.1201/b15895} {\emph {\bibinfo {title}
  {Fundamental {{Number Theory}} with {{Applications}}}}},\ \bibinfo {edition}
  {zeroth}\ ed.\ (\bibinfo  {publisher} {{Chapman and Hall/CRC}},\ \bibinfo
  {year} {2008})\BibitemShut {NoStop}%
\bibitem [{\citenamefont {Storjohann}(2000)}]{storjohann2000}%
  \BibitemOpen
  \bibfield  {author} {\bibinfo {author} {\bibfnamefont {A.}~\bibnamefont
  {Storjohann}},\ }\href {https://doi.org/10.3929/ethz-a-004141007} {\bibfield
  {journal} {\bibinfo  {journal} {Dissertation, Swiss Federal Institute of
  Technology, Zurich}\ } (\bibinfo {year} {2000})}\BibitemShut {NoStop}%
\bibitem [{\citenamefont {Noh}\ \emph {et~al.}(2022)\citenamefont {Noh},
  \citenamefont {Chamberland},\ and\ \citenamefont {Brand{\~a}o}}]{noh2022low}%
  \BibitemOpen
  \bibfield  {author} {\bibinfo {author} {\bibfnamefont {K.}~\bibnamefont
  {Noh}}, \bibinfo {author} {\bibfnamefont {C.}~\bibnamefont {Chamberland}},\
  and\ \bibinfo {author} {\bibfnamefont {F.~G.}\ \bibnamefont {Brand{\~a}o}},\
  }\href@noop {} {\bibfield  {journal} {\bibinfo  {journal} {PRX Quantum}\
  }\textbf {\bibinfo {volume} {3}},\ \bibinfo {pages} {010315} (\bibinfo {year}
  {2022})}\BibitemShut {NoStop}%
\bibitem [{\citenamefont {Baragiola}(2022)}]{privateben}%
  \BibitemOpen
  \bibfield  {author} {\bibinfo {author} {\bibfnamefont {B.~Q.}\ \bibnamefont
  {Baragiola}},\ }\href@noop {} {}\bibinfo {howpublished} {private
  communication} (\bibinfo {year} {2022})\BibitemShut {NoStop}%
\bibitem [{Note4()}]{Note4}%
  \BibitemOpen
  \bibinfo {note} {Gaussian operations parameterized by irrational symplectic
  operations cannot in general be simulated with our method. We refer to our
  previous work~\cite {calcluth2022} which demonstrates that when the
  symplectic matrix is irrational, the wavefunction of the transformed state
  corresponds to a periodic distribution which cannot be analytically reduced.
  Measuring in the position basis of a state which has been transformed by a
  general irrational symplectic matrix will have a PDF which will give random
  integer combinations of irrational numbers. Except for specific choices of
  irrational symplectic matrices, the measurement values will be randomly
  selected from a set dense on the real number line.}\BibitemShut {Stop}%
\bibitem [{\citenamefont {Croom}(2016)}]{croom2016}%
  \BibitemOpen
  \bibfield  {author} {\bibinfo {author} {\bibfnamefont {F.~H.}\ \bibnamefont
  {Croom}},\ }\href@noop {} {\emph {\bibinfo {title} {Principles of
  Topology}}},\ \bibinfo {edition} {dover edition}\ ed.\ (\bibinfo  {publisher}
  {{Dover Publications, Inc}},\ \bibinfo {address} {{Mineola, New York}},\
  \bibinfo {year} {2016})\BibitemShut {NoStop}%
\bibitem [{sta(2022{\natexlab{b}})}]{stackexchangedensity}%
  \BibitemOpen
  \href@noop {} {\bibinfo {title} {Is the symplectic group over the rationals
  dense on the symplectic group over the reals?}} (\bibinfo {year}
  {2022}{\natexlab{b}}),\ \bibinfo {note} {{Mathematics Stack Exchange}.
  Available at \url{https://math.stackexchange.com/q/4510323/} (accessed:
  2022-08-18)}\BibitemShut {NoStop}%
\bibitem [{\citenamefont {Artin}(1988)}]{artin1988}%
  \BibitemOpen
  \bibfield  {author} {\bibinfo {author} {\bibfnamefont {E.}~\bibnamefont
  {Artin}},\ }\href@noop {} {\emph {\bibinfo {title} {Geometric Algebra}}},\
  Wiley Classics Library\ (\bibinfo  {publisher} {{J. Wiley}},\ \bibinfo
  {address} {{New York}},\ \bibinfo {year} {1988})\BibitemShut {NoStop}%
\bibitem [{\citenamefont {Weyl}(1946)}]{weyl1946}%
  \BibitemOpen
  \bibfield  {author} {\bibinfo {author} {\bibfnamefont {H.}~\bibnamefont
  {Weyl}},\ }\href@noop {} {\emph {\bibinfo {title} {The Classical Groups:
  Their Invariants and Representations}}},\ \bibinfo {edition} {2nd}\ ed.,\
  Princeton Landmarks in Mathematics and Physics {{Mathematics}}\ (\bibinfo
  {publisher} {{Princeton University Press}},\ \bibinfo {address} {{Princeton,
  N.J. Chichester}},\ \bibinfo {year} {1946})\BibitemShut {NoStop}%
\bibitem [{\citenamefont {Pedoe}(1988)}]{pedoe1988}%
  \BibitemOpen
  \bibfield  {author} {\bibinfo {author} {\bibfnamefont {D.}~\bibnamefont
  {Pedoe}},\ }\href@noop {} {\emph {\bibinfo {title} {Geometry, a Comprehensive
  Course}}}\ (\bibinfo  {publisher} {{Dover Publications}},\ \bibinfo {address}
  {{New York}},\ \bibinfo {year} {1988})\BibitemShut {NoStop}%
\bibitem [{\citenamefont {Trench}(2003)}]{trench2003}%
  \BibitemOpen
  \bibfield  {author} {\bibinfo {author} {\bibfnamefont {W.~F.}\ \bibnamefont
  {Trench}},\ }\href@noop {} {\emph {\bibinfo {title} {Introduction to real
  analysis}}}\ (\bibinfo  {publisher} {Prentice Hall/Pearson Education},\
  \bibinfo {address} {Upper Saddle River, N.J},\ \bibinfo {year}
  {2003})\BibitemShut {NoStop}%
\bibitem [{\citenamefont {Greub}(1975)}]{greub1975}%
  \BibitemOpen
  \bibfield  {author} {\bibinfo {author} {\bibfnamefont {W.}~\bibnamefont
  {Greub}},\ }\href {https://doi.org/10.1007/978-1-4684-9446-4} {\emph
  {\bibinfo {title} {Linear {Algebra}}}},\ \bibinfo {series} {Graduate {Texts}
  in {Mathematics}}, Vol.~\bibinfo {volume} {23}\ (\bibinfo  {publisher}
  {Springer},\ \bibinfo {address} {New York},\ \bibinfo {year}
  {1975})\BibitemShut {NoStop}%
\bibitem [{\citenamefont {Roman}(2007)}]{roman2007}%
  \BibitemOpen
  \bibfield  {author} {\bibinfo {author} {\bibfnamefont {S.}~\bibnamefont
  {Roman}},\ }\href@noop {} {\emph {\bibinfo {title} {Advanced Linear
  Algebra}}},\ \bibinfo {edition} {3rd}\ ed.,\ \bibinfo {series} {Graduate
  Texts in Mathematics}\ No.\ \bibinfo {number} {135}\ (\bibinfo  {publisher}
  {{Springer}},\ \bibinfo {address} {{New York}},\ \bibinfo {year}
  {2007})\BibitemShut {NoStop}%
\bibitem [{\citenamefont {Golub}\ and\ \citenamefont {Loan}(1996)}]{golub1996}%
  \BibitemOpen
  \bibfield  {author} {\bibinfo {author} {\bibfnamefont {G.~H.}\ \bibnamefont
  {Golub}}\ and\ \bibinfo {author} {\bibfnamefont {C.~F.~V.}\ \bibnamefont
  {Loan}},\ }\href@noop {} {\emph {\bibinfo {title} {Matrix
  {{Computations}}}}},\ \bibinfo {edition} {3rd}\ ed.\ (\bibinfo  {publisher}
  {{The John Hopkins University Press}},\ \bibinfo {year} {1996})\BibitemShut
  {NoStop}%
\bibitem [{\citenamefont {Greville}(1966)}]{greville1966}%
  \BibitemOpen
  \bibfield  {author} {\bibinfo {author} {\bibfnamefont {T.~N.~E.}\
  \bibnamefont {Greville}},\ }\href {https://doi.org/10.1137/1008107}
  {\bibfield  {journal} {\bibinfo  {journal} {SIAM Review}\ }\textbf {\bibinfo
  {volume} {8}},\ \bibinfo {pages} {518} (\bibinfo {year} {1966})}\BibitemShut
  {NoStop}%
\bibitem [{\citenamefont {Arvind}\ \emph {et~al.}(1995)\citenamefont {Arvind},
  \citenamefont {Dutta}, \citenamefont {Mukunda},\ and\ \citenamefont
  {Simon}}]{arvind1995}%
  \BibitemOpen
  \bibfield  {author} {\bibinfo {author} {\bibnamefont {Arvind}}, \bibinfo
  {author} {\bibfnamefont {B.}~\bibnamefont {Dutta}}, \bibinfo {author}
  {\bibfnamefont {N.}~\bibnamefont {Mukunda}},\ and\ \bibinfo {author}
  {\bibfnamefont {R.}~\bibnamefont {Simon}},\ }\href@noop {} {\bibfield
  {journal} {\bibinfo  {journal} {Pramana J. Phys.}\ }\textbf {\bibinfo
  {volume} {45}},\ \bibinfo {pages} {441} (\bibinfo {year} {1995})}\BibitemShut
  {NoStop}%
\bibitem [{\citenamefont {Dummit}\ and\ \citenamefont
  {Foote}(1991)}]{dummit1991}%
  \BibitemOpen
  \bibfield  {author} {\bibinfo {author} {\bibfnamefont {D.~S.}\ \bibnamefont
  {Dummit}}\ and\ \bibinfo {author} {\bibfnamefont {R.~M.}\ \bibnamefont
  {Foote}},\ }\href@noop {} {\emph {\bibinfo {title} {Abstract algebra}}},\
  Vol.\ \bibinfo {volume} {1999}\ (\bibinfo  {publisher} {Prentice Hall
  Englewood Cliffs, NJ},\ \bibinfo {year} {1991})\BibitemShut {NoStop}%
\bibitem [{\citenamefont {Bermejo-Vega}(2016)}]{Juani-thesis}%
  \BibitemOpen
  \bibfield  {author} {\bibinfo {author} {\bibfnamefont {J.}~\bibnamefont
  {Bermejo-Vega}},\ }\href {https://arxiv.org/abs/1611.09274} {\bibfield
  {journal} {\bibinfo  {journal} {PhD Thesis, Technische Universit\"at
  M\"unchen Max-Planck-Institut f\"ur Quantenoptik}\ } (\bibinfo {year}
  {2016})},\ \bibinfo {note} {arXiv:1611.09274}\BibitemShut {NoStop}%
\bibitem [{\citenamefont {Bermejo-Vega}\ \emph {et~al.}(2016)\citenamefont
  {Bermejo-Vega}, \citenamefont {Lin},\ and\ \citenamefont {Van~den
  Nest}}]{bermejo-vega2016}%
  \BibitemOpen
  \bibfield  {author} {\bibinfo {author} {\bibfnamefont {J.}~\bibnamefont
  {Bermejo-Vega}}, \bibinfo {author} {\bibfnamefont {Y.}~\bibnamefont {Lin}},\
  and\ \bibinfo {author} {\bibfnamefont {M.}~\bibnamefont {Van~den Nest}},\
  }\href@noop {} {\bibfield  {journal} {\bibinfo  {journal} {Quantum
  Information and Computation}\ }\textbf {\bibinfo {volume} {16}},\ \bibinfo
  {pages} {0361} (\bibinfo {year} {2016})}\BibitemShut {NoStop}%
\end{thebibliography}%

\end{document}